\documentclass[fleqn,usenatbib]{mnras}
 
\usepackage{graphicx} 
\usepackage{txfonts} 
\usepackage{natbib}  
\usepackage{multirow}
\usepackage{multicol}
\title[Chemical composition of 1G stars in GCs]{Constraining the original composition of the gas forming first-generation stars in globular clusters}

\author[M.\,V.\,Legnardi et al.] 
       {M.\,V.\,Legnardi$^{1}$\thanks{E-mail: mariavittoria.legnardi@phd.unipd.it},
       A.\,P.\,Milone$^{1,2}$,
       L.\,Armillotta$^{3}$,
       A.\,F.\,Marino$^{4}$,
       G.\,Cordoni$^{1}$,
       A.\,Renzini$^{2}$,
       \newauthor E.\,Vesperini$^{5}$,
       F.\,D'Antona$^{6}$,
       M.\,McKenzie$^{7,8}$,
       D.\,Yong$^{8}$,
       E.\,Dondoglio$^{1}$,
       E.\,P.\,Lagioia$^{1}$,
       \newauthor M.\,Carlos$^{1}$,
       M.\,Tailo$^{9}$,
       S.\,Jang$^{1}$,
       and A.\,Mohandasan$^{1}$
\\ 
$^{1}$ Dipartimento di Fisica e Astronomia ``Galileo Galilei'', Univ. di Padova, Vicolo dell'Osservatorio 3, Padova, IT-35122\\
$^{2}$ Istituto Nazionale di Astrofisica - Osservatorio Astronomico di Padova, Vicolo dell'Osservatorio 5, IT-35122, Padua, Italy \\
$^{3}$ Department of Astrophysical Sciences, Princeton University, Princeton, NJ 08544, USA \\
$^{4}$ Istituto Nazionale di Astrofisica - Osservatorio Astrofisico di Arcetri, Largo Enrico Fermi, 5, Firenze, IT-50125 \\
$^{5}$ Department of Astronomy, Indiana University, Bloomington, IN 47401, USA\\
$^{6}$ INAF - Osservatorio Astronomico di Roma, Via Frascati 33, I-00040, Monte Porzio Catone, Roma, Italy \\
$^{7}$ ICRAR, M468, The University of Western Australia, 35 Stirling Highway, Crawley, WA 6009, Australia \\
$^{8}$ Research School of Astronomy \& Astrophysics, Australian National University, Canberra, ACT 2611, Australia \\
$^{9}$ Dipartimento di Fisica e Astronomia Augusto Righi, Universit\`{a} degli Studi di Bologna, Via Gobetti 93/2, I-40129 Bologna, Italy \\
} 
\begin{document} 
\label{firstpage}
\date{Accepted 2022 March 11. Received 2022 March 10; in original form 2022 February 4}
 
\pagerange{\pageref{firstpage}--\pageref{lastpage}} \pubyear{2022} 
 
\maketitle 

\begin{abstract}
Disentangling distinct stellar populations along the red-giant branches (RGBs) of Globular Clusters (GCs) is possible by using the pseudo two-color diagram dubbed chromosome map (ChM). One of the most intriguing findings is that the so-called first-generation (1G) stars, characterized by the same chemical composition of their natal cloud, exhibit extended sequences in the ChM. Unresolved binaries and internal variations in helium or metallicity have been suggested to explain this phenomenon. Here, we derive high-precision {\it Hubble Space Telescope} photometry of the GCs NGC\,6362 and NGC\,6838 and build their ChMs. We find that both 1G RGB and main-sequence (MS) stars exhibit wider ChM sequences than those of second-generation (2G). The evidence of this feature even among unevolved 1G MS stars indicates that chemical inhomogeneities are imprinted in the original gas. We introduce a pseudo two-magnitude diagram to distinguish between helium and metallicity, and demonstrate that star-to-star metallicity variations are responsible for the extended 1G sequence. Conversely, binaries provide a minor contribution to the phenomenon. We estimate that the metallicity variations within 1G stars of 55 GCs range from less than [Fe/H]$\sim$0.05 to $\sim$0.30 and mildly correlate with cluster mass. We exploit these findings to constrain the formation scenarios of multiple populations showing that they are qualitatively consistent with the occurrence of multiple generations. In contrast, the fact that 2G stars have more homogeneous iron content than the 1G challenges the scenarios based on accretion of material processed in massive 1G stars onto existing protostars.
\end{abstract} 
 
\begin{keywords} 
  globular clusters: general, stars: population II, stars: abundances, techniques: photometry.
\end{keywords}

\section {Introduction}
\label{sec:intro}
Globular Clusters (GCs) are thought to form out of high-density regions within supergiant molecular clouds that trace back to earliest epochs \citep[e.g.,][]{harris1994}. While these objects are of fundamental importance for a range of astrophysical studies, the origin of such compact and massive agglomerates of stars is still strongly debated. Conversely, it is now a well-established fact that GCs host multiple stellar populations with different chemical composition. A considerable number of works has provided strong evidence that these stellar systems harbor at least two main populations of stars, commonly dubbed first- (1G) and second-generation (2G).  The latter includes stars that, being enhanced in He, N, and Na and depleted in C and O, exhibit abundance patterns distinctive of GC stars. On the contrary, 1G stars are characterized by an ‘original’ chemical composition that retains memories of their natal cloud chemistry \citep[e.g.,][]{kraft1994a, carretta2009a, marino2019a}. In light of this feature and since Galactic GCs are among the most ancient objects of the Galaxy, studying 1G stars provides the unique opportunity to trace the original chemical composition of the clouds where GCs formed in the early Universe. 

A powerful tool to infer the relative chemical composition of GC stars is the ‘Chromosome Map’ \citep[ChM;][]{milone2015a}, a pseudo two-color diagram, constructed with appropriate photometric bands of the {\it Hubble Space Telescope} ({\it HST}), which is particularly sensitive to the chemistry of the distinct stellar populations thus maximizing the separation between them. On the ChM reference frame, $\Delta_{\rm {\it C}\,F275W, F336W, F438W}$ vs.\,$\Delta_{\rm F275W, F814W}$\footnote{A brief description of the procedure used to build the ChM can be found in Section~\ref{sec:MP_ChM}. In addition, see \citet{milone2015a,milone2017a} for a comprehensive definition of the $\Delta_{\rm {\it C}\,F275W, F336W, F438W}$ and $\Delta_{\rm F275W, F814W}$ pseudo-colors.}, the 2G sequence is elongated towards large $\Delta_{\rm {\it C}\,F275W, F336W, F438W}$ and low $\Delta_{\rm F275W, F814W}$ values, whereas 1G stars lay in proximity of the origin. As pointed out by \cite{milone2017a}, the $\Delta_{\rm F275W, F814W}$ color and $\Delta_{\rm {\it C}\,F275W, F336W, F438W}$ pseudo-color distributions of both 1G and 2G stars are wider than photometric errors (including errors associated to differential reddening corrections), thus demonstrating that neither of the two components is consistent with simple stellar populations. This phenomenon can be naturally explained for the 2G component using the dependence of $\Delta_{\rm F275W, F814W}$ and $\Delta_{\rm {\it C}\,F275W, F336W, F438W}$ pseudo-colors on light-element abundances. As illustrated in details by \citet[see Section 4]{milone2018a}, indeed, the 2G color spread can be reproduced by a combination of variations in the N and He abundances, as expected for 2G stars characterized by a chemical composition produced by various degrees of CNO processing. The extension of the 1G sequence, instead, has revealed to be a more puzzling result.

Given that the position of stars along the ChM x-axis of monometallic GCs is predominantly affected by their helium abundance, \cite{milone2018a} tentatively attributed the 1G color spread to a pure helium variation without any appreciable enrichment in other light elements, like N. Inhomogeneities in the helium content of the gas forming 1G stars may arise in regions of the Universe where the baryon-to-photon ratio was considerably enhanced, possibly by many orders of magnitude \citep[e.g.,][]{arbey2020}. Conversely, according to basic stellar nucleosynthesis, reproducing significant helium enhancements (even of the order of $\delta Y_{\rm 1G}\sim0.1$) not accompanied by appreciable nitrogen variations, is extremely difficult and all the scenarios considered by \citet[][see their discussion in Section~8]{milone2018a} are unable to reproduce observations. In addition, the analysis of M\,3 horizontal-branch (HB) stars by \cite{tailo2019a} demonstrates that pure helium variations among 1G stars lead to inconsistent properties of HB stars thus making this scenario even more implausible. 

An alternative hypothesis was suggested first by \citet[]{dantona2016a} based on theoretical arguments and then by spectroscopy of 1G stars by \cite{marino2019a, marino2019b}. According to these authors, variations in iron abundances could mimic the effect on the 1G ChM color distribution caused by a spread in helium. \citet[]{marino2019b} analysed the chemical abundances of 18 red-giant branch (RGB) stars belonging to the 1G of NGC\,3201 finding a spread in the overall metallicity of the order of $\sim0.1$ dex. The fact that GC stars are not chemically homogeneous is additionally supported by the pioneering work by \citet[][]{yong2013a}, based on high-precision differential abundances of 1G and 2G stars in the GC NGC\,6752. Iron variations in the gas forming 1G stars might have been inherited from the interstellar medium (ISM) out of which proto-cluster molecular clouds formed. In turn, chemical inhomogeneities in the ISM might result from incomplete mixing of supernova ejecta with the ISM \citep[e.g.,][]{krumholz2018,bailin2021,wirth2021}. Alternatively, iron variations might be the product of stellar feedback inside the cloud itself \citep[e.g.,][]{mckenzie2021}.

Among the investigated targets, Marino and collaborators identified three binary candidates as the stars with the lowest $\Delta_{\rm F275W,F814W}$ values, thus suggesting that binarity could be responsible for the elongation of the 1G component. The authors explored deeply this possibility demonstrating that a large number of binaries is required to account for the observed 1G color spread. This finding allows to conclude that binaries provide only a minor contribution to the elongation of the 1G sequence \citep[see also][]{kamann2020a, martins2020}. 

More recently, extended sequences of 1G stars have been also detected along the red HB of 12 Galactic GCs \citep[][]{dondoglio2021a}, thus demonstrating that this phenomenon is not a peculiarity of RGB stars. The spectro-photometric analysis by Dondoglio and collaborators have shown that the extended red-HB sequences are consistent with either star-to-star iron variations or with an internal helium spread, in close analogy with what has been observed along the RGB ChM. 

With the aim of shedding new light on this intriguing phenomenon, in this work we combine multi-band {\it HST} photometry and synthetic spectra analysis techniques to investigate for the first time chemical variations among unevolved main-sequence (MS) stars of the Galactic GCs NGC\,6362 and NGC\,6838. We choose to analyze these two clusters because the ChM of their RGB stars display an elongated color sequence that is not consistent with observational errors alone \citep[see Figure 3 and 4 of][]{milone2017a}. Moreover, they display a quite simple multiple population pattern with only two distinct groups of 1G and 2G stars that reveal to be well separated at all evolutionary stages, both in ChMs and other appropriate photometric diagrams. For a general overview of the two GCs we listed their main parameters in Table~\ref{GC_parameters}.
 
The paper is organized as follows. Section~\ref{sec:data} presents the photometric dataset for NGC\,6362 and NGC\,6838 and the procedure employed to reduce it. In Section~\ref{sec:MP_ChM} we investigate multiple stellar populations and derive the ChM of MS stars for both clusters. A detailed analysis of the 1G color spread observed on the ChM is developed in Section~\ref{sec:1g_extended} and Section~\ref{sec:1g_SGB} by using MS and sub-giant branch (SGB) stars, respectively. In Section~\ref{sec:correlations} we extend our considerations to a sample of 55 Galactic GCs calculating the internal iron variations within the 1G and searching for correlations between this quantity and the main host GC parameters. Finally, in Section~\ref{sec:conclusion} we summarize the main results of this work and discuss physical reasons that could explain the presence of iron inhomogeneities among 1G stars. We also speculate about possible implications on the formation scenarios of 2G stars within GCs.    

\begin{table*}
    \caption{Main parameters of NGC\,6362 and NGC\,6838. Metallicities are derived from high-resolution spectroscopy by \citet{massari2017a} and \citet{carretta2009b}. The values of distance modulus, age, and reddening are taken from \citet{dotter2010}, whereas absolute visual magnitudes and cluster masses are provided by the 2010 version of the \citet{harris1996} catalogue and \citet{baumgardt2018}, respectively.}
    \centering
    \begin{tabular}{ccccccc}
    \hline\hline
         Cluster ID & [Fe/H] & ({\it m}$-${\it M})$_{\rm V}$ & Age & E(B$-$V) & {\it M}$_{V}$ & Mass \\
         & & (mag) & (Gyr) & (mag) & (mag) & (M$_{\odot}$) \\
         \hline
         NGC\,6362 & $-$1.07$\pm$0.01 & 14.55 & 12.5$\pm$0.50 & 0.07 & $-$6.95 & 1.47$\pm$0.04 $\times$10$^{5}$ \\
         NGC\,6838 & $-$0.82$\pm$0.02 & 13.40 & 12.5$\pm$0.75 & 0.22 & $-$5.61 & 4.91$\pm$0.47 $\times$10$^{4}$  \\
         \hline\hline
    \end{tabular}
    \label{GC_parameters}
\end{table*}
\begin{table*}
\caption{Summary information about the archive images of NGC\,6362 and NGC\,6838 used in this work.}
    \centering
    \begin{tabular}{cccccc}
    \hline \hline
    \\
         DATE & N $\times$ EXPTIME & INSTRUMENT & FILTER & PROGRAM & PI   \\
         \hline 
         & & {\large NGC\,6362} & & & \\
         \\
         2014 March 30 & $2\times720$\,s & UVIS/WFC3 & F275W & 13297 & G. Piotto \\
         2014 July 01 & $2\times829$\,s & UVIS/WFC3 & F275W & 13297 & G. Piotto \\
         2010 August 13 & $368\rm \,s+4\times450$\,s & UVIS/WFC3 & F336W & 12008 & A. Kong \\
         2014 March 29 & 323\,s & UVIS/WFC3 & F336W & 13297 & G. Piotto \\
         2014 March 30 & 323\,s & UVIS/WFC3 & F336W & 13297 & G. Piotto \\
         2014 July 01 & $2\times323$\,s & UVIS/WFC3 & F336W & 13297 & G. Piotto \\
         2014 March 29 & 68\,s & UVIS/WFC3 & F438W & 13297 & G. Piotto \\
         2014 July 01 & 67\,s & UVIS/WFC3 & F438W & 13297 & G. Piotto \\
         2006 May 30 & $10 \rm \,s+4\times130$\,s & WFC/ACS & F606W & 10775 & A. Sarajedini \\
         2011 March 30 & $140\rm\,s+145\rm\,s$ & WFC/ACS & F625W & 12008 & A. Kong \\
         2011 March 30 & $750 \rm\,s+766 \rm\,s$ & WFC/ACS & F658N & 12008 & A. Kong \\
         2006 May 30 & $10 \rm\,s+4\times150$\,s & WFC/ACS & F814W & 10775 & A. Sarajedini \\
         \\
          & & {\large NGC\,6838} & & & \\
         \\
         2013 October 23 & $2\times792$\,s & UVIS/WFC3 & F275W & 13297 & G. Piotto \\
         2014 May 03 & $2\times750$\,s & UVIS/WFC3 & F275W & 13297 & G. Piotto \\
         2013 October 23 & $2\times303$\,s & UVIS/WFC3 & F336W & 13297 & G. Piotto \\
         2014 May 03 & $2\times303$\,s & UVIS/WFC3 & F336W & 13297 & G. Piotto \\
         2013 October 23 & 65\,s & UVIS/WFC3 & F438W & 13297 & G. Piotto \\
         2014 May 03 & 65\,s & UVIS/WFC3 & F438W & 13297 & G. Piotto \\
         2006 May 12 & $4\rm \,s+4\times75$\,s & WFC/ACS & F606W & 10775 & A. Sarajedini \\
         2013 August 20 & $5\times500\rm \,s+3\times466\rm \,s+2\times459\rm \,s$ & WFC/ACS & F606W & 12932 & F. Ferraro \\ 
         2006 May 12 & $4\rm \,s+4\times80$\,s & WFC/ACS & F814W & 10775 & A. Sarajedini \\
         2013 August 20 & $440\rm \,s+3\times357\rm \,s+5\times337\rm \,s$ & WFC/ACS & F814W & 12932 & F. Ferraro \\
         2020 December 12 & $35\rm \,s+4\times337\rm \,s$ & WFC/ACS & F814W & 16298 & M. Libralato \\
         
         \hline \hline
    \end{tabular}
    \label{dataset_table}
\end{table*}

\section{Data and Data reduction}
\label{sec:data}
In this work the astro-photometric catalogs of the Galactic GCs NGC\,6362 and NGC\,6838 were obtained starting from the archive images taken with the Wide Field Camera of the Advanced Camera for Surveys (WFC/ACS) and the Ultraviolet and Visual Channel of the Wide Field Camera 3 (UVIS/WFC3) on board {\it HST}. Specifically, to investigate the 1G color spread in NGC\,6838 we used exposures collected through 5 filters, namely F275W, F336W, F438W, F606W, and F814W. A similar dataset has been exploited for NGC\,6362 with the only exception that also images in the F625W and F658N filters of WFC/ACS were available. The main properties of the complete set of exposures used in this work are summarized in Table~\ref{dataset_table}. 

Accurate measurements of stellar position and magnitudes have been derived by using the computer program KS2 developed by Jay Anderson as an evolution of \texttt{kitchen\_sync} \citep{anderson2008a}. Briefly, KS2 performs different iterations to find and measure stars: it first identifies only the brightest stars and then, after subtracting them, it looks for progressively fainter stars that satisfy a set of different criteria based on the distance, on the peak position or on the quality fit parameter. This program offers the possibility to measure stars with distinct methods optimized to provide the best possible photometry according to their luminosities \citep[see, e.g.,][for a detailed description of how KS2 works]{sabbi2016, bellini2017, nardiello2018}. Specifically, the positions and magnitudes of relatively-bright sources, having enough flux to be detected in single images, are measured by fitting the appropriate PSF model in each exposure independently and then averaged together to derive the best estimates. Since faint stars do not produce a significant peak in all the exposures this approach is not suited to measure them making necessary to combine the information from all the images to derive their positions and magnitudes. 

KS2 also provides various diagnostics to test whether the photometric and astrometric measurements are reliable or not. Following the procedure described in \cite{milone2009} and \cite{bedin2009a}, we exploited these parameters to select a sample of relatively isolated stars well-fitted by the PSF. 

The next step was to calibrate photometry into the Vega system. To do that we exploited the procedure by \cite{nardiello2018} using the updated zero points retrieved from the Space Telescope Science Institute webpage\footnote{\url{https://acszeropoints.stsci.
edu} and \url{https://www.stsci.edu/hst/instrumentation/
wfc3/data-analysis/photometric-calibration} for ACS/WFC and UVIS/WFC3, respectively}. Finally, magnitudes have been corrected for the effects of differential reddening following the recipe introduced by \citet[see their Section 3 for details on the procedure]{milone2012redd}. 

\subsection{Artificial stars}
To estimate the photometric uncertainties associated to our measurements and generate simulated diagrams we performed artificial-star (AS) tests extending the procedure introduced by \cite{anderson2008a} to NGC\,6362 and NGC\,6838.

Briefly, we first generated a list of coordinates and magnitudes of 99,999 stars with a spatial distribution along the field of view resembling the one of cluster stars. The magnitudes of ASs were derived starting from a set of fiducial lines obtained from the observed color-magnitude diagrams (CMDs). 

We reduced ASs by using once again KS2. In particular, the program derives for ASs the same parameters used to evaluate the astrometric and photometric quality of real stars. To select a sample of relatively-isolated ASs well-fitted by the PSF we applied the same stringent criteria obtained from the reduction of real stars. 
\begin{figure*} 
\begin{center} 
    \centering
    \includegraphics[scale=.9]{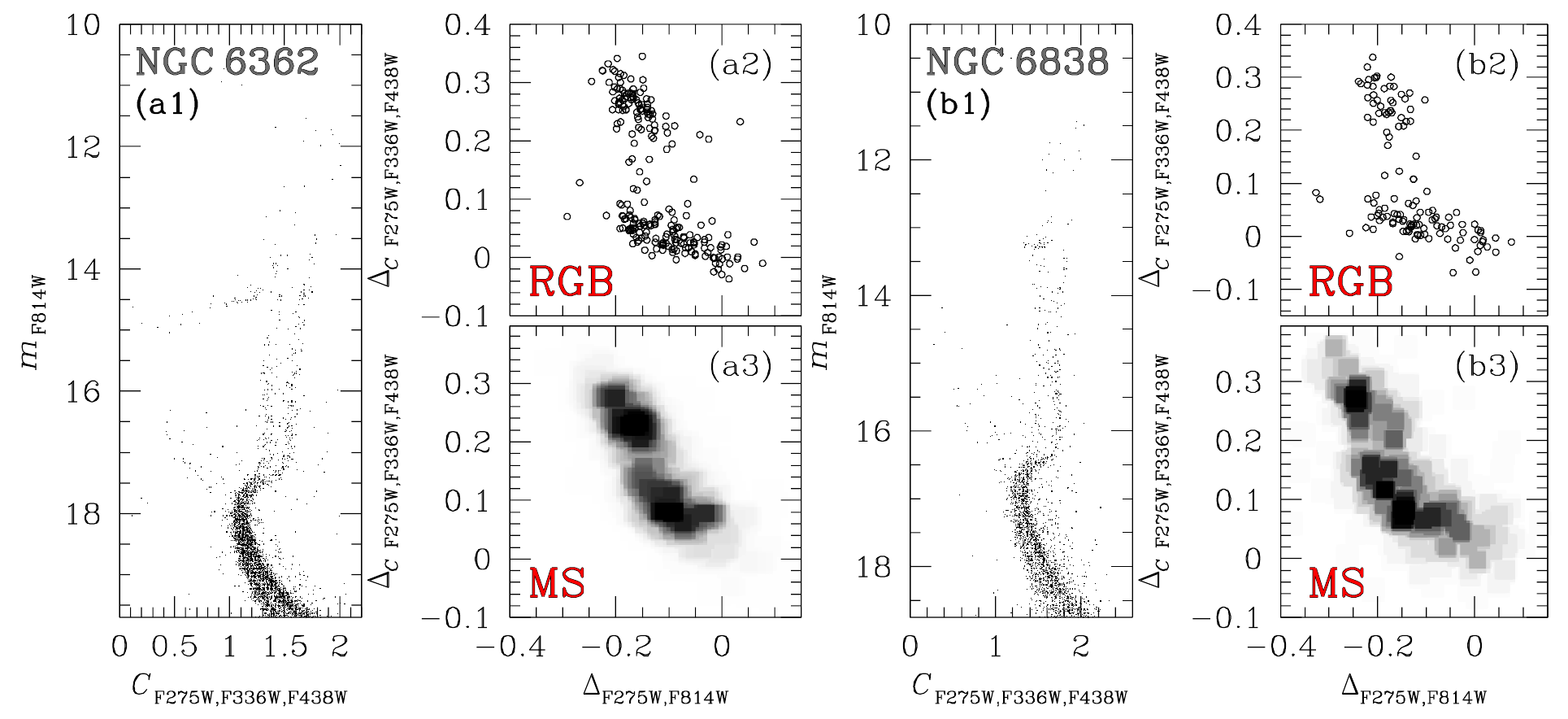}
    \caption{Panels a1 and b1 show the $m_{\rm F814W}$ vs.\,$C_{\rm F275W,F336W,F438W}$ CMDs for NGC\,6362 and NGC\,6838, respectively. This diagram has been exploited in combination with the $m_{\rm F814W}$ vs.\,$m_{\rm F275W}-m_{\rm F814W}$ CMD to derive the $\Delta_{\rm F275W,F814W}$ and the $\Delta_{\rm {\it C}\,F275W,F336W,F438W}$ pseudo-colors used to build the ChMs of MS and RGB stars. The ChMs of RGB stars are shown in panels a2-b2, whereas the Hess diagrams of the MS ChMs are plotted in panels a3-b3.} \label{fig:plot_cmd}
\end{center}
\end{figure*} 
\begin{figure*} 
  \centering
   \includegraphics[height=9.cm,trim={3.25cm 1.25cm 0.4cm 9.6cm},clip]{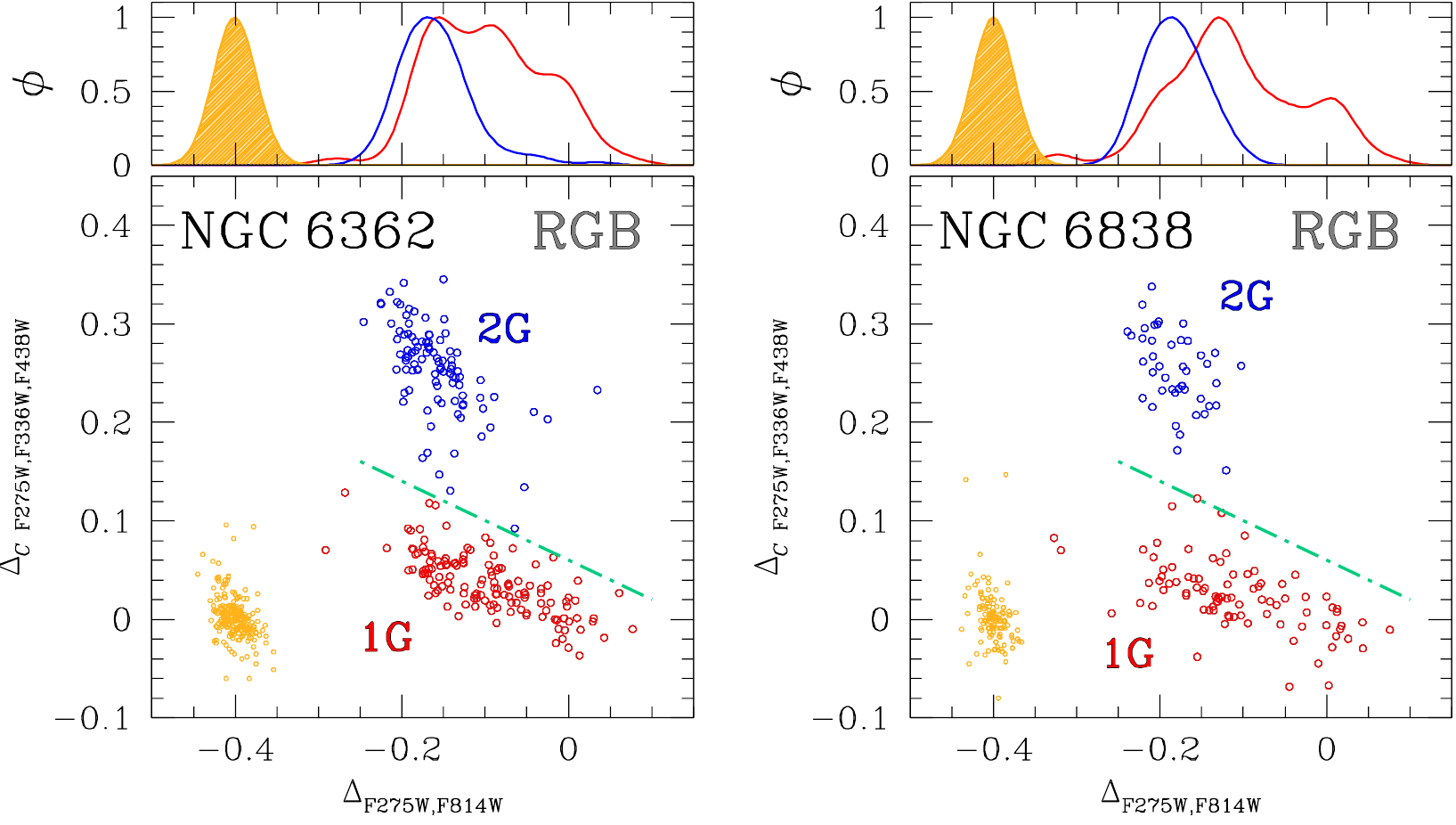}
   \includegraphics[height=9.cm,trim={3.25cm 1.25cm 0.4cm 9.6cm},clip]{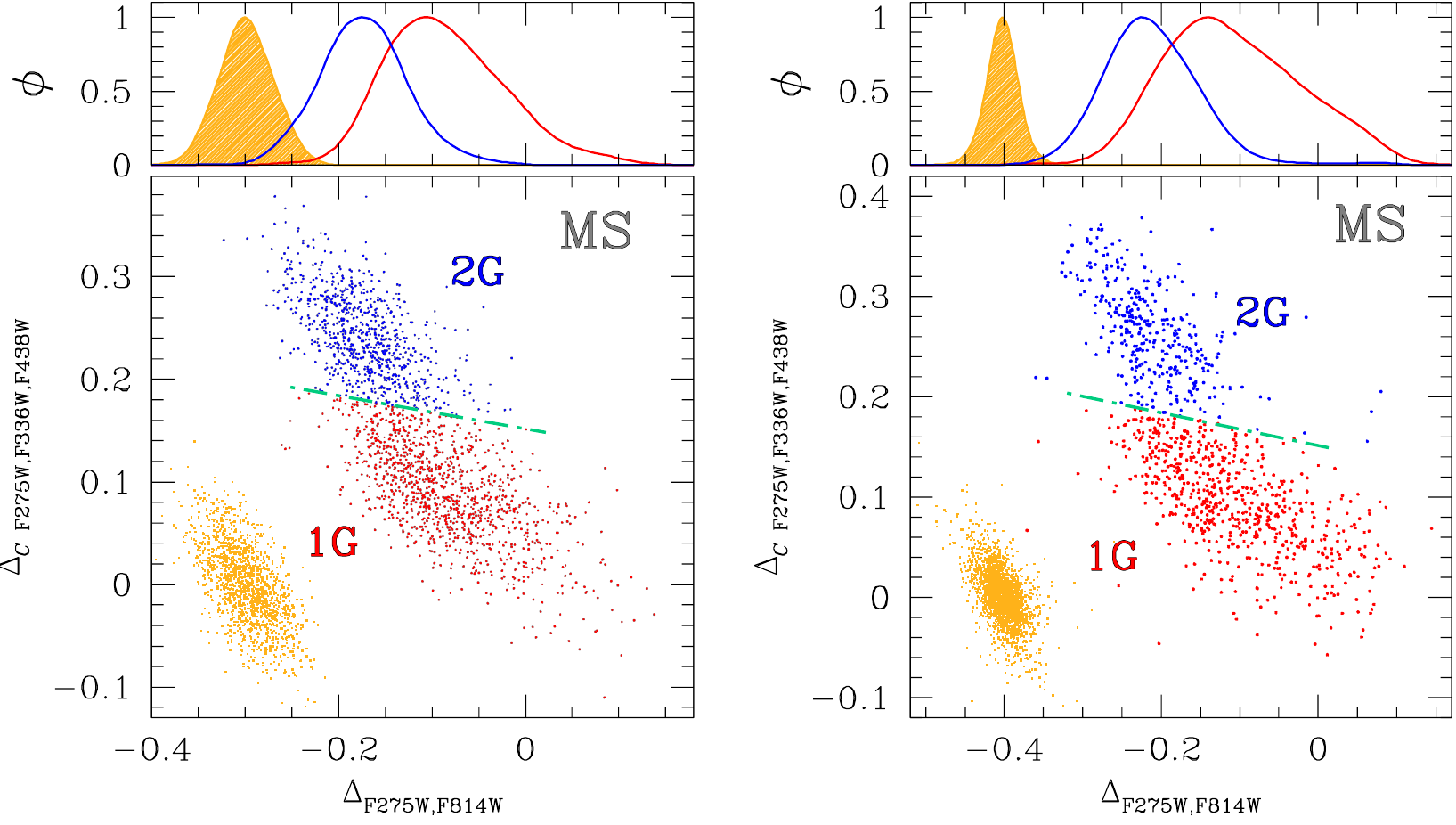}
  \caption{ChM of RGB (upper panels) and MS (lower panels) stars in NGC\,6362 (left) and NGC\,6838 (right). 1G and 2G stars are colored red and blue, respectively, while the error distribution is represented by orange points on the bottom-left corner. The kernel-density distributions of 1G, 2G stars, and observational errors are plotted on the top panel of each figure.} 
  \label{fig:ChMs}
\end{figure*} 

\section{Multiple populations in NGC 6362 and NGC 6838}
\label{sec:MP_ChM}
A visual inspection at the $m_{\rm F814W}$ vs.\,$C_{\rm F275W,F336W,F438W}$ diagrams of NGC\,6362 and NGC\,6838 illustrated in panels a1-b1 of Figure~\ref{fig:plot_cmd}, respectively, immediately reveals that both clusters exhibit two distinct sequences that run almost parallel from the bottom of the MS to the upper part of the RGB. These sequences correspond to stellar populations with different abundances in light elements. Indeed, the $C_{\rm F275W,F336W,F438W}$ pseudo-color, encompassing the absorption features of the OH, NH, CH, and CN molecules, allows us to separate stellar populations with different content of C, N, and O.

To better investigate stellar populations, we show in panels a2-b2 of Figure~\ref{fig:plot_cmd} the ChMs of RGB  stars for NGC\,6362 and NGC\,6838, respectively \citep[][]{milone2017a}. Similarly, panels  a3-b3 of Figure~\ref{fig:plot_cmd} reproduce the Hess diagram of the ChM for MS stars. These diagrams were derived adapting the method by \cite{milone2015a, milone2017a} to MS stars. In the case of NGC\,6362, for example, we first selected stars with $18.75\le m_{\rm F814W} \le19.75$ in the $m_{\rm F814W}$ vs.\,$C_{\rm F275W,F336W,F438W}$ and $m_{\rm F814W}$ vs.\,$m_{\rm F275W}-m_{\rm F814W}$ CMDs and divided this sample of MS stars into magnitude bins of size $0.2$\,mag. For each bin, we calculated the 4$^{\rm th}$ and the 96$^{\rm th}$ percentile of the $m_{\rm F275W}-m_{\rm F814W}$ color and $C_{\rm F275W,F336W,F438W}$ pseudo-color distribution. We then interpolated these values with the mean $m_{\rm F814W}$ magnitudes to infer the blue and red boundaries of the selected MS stars. Finally, to obtain the $\Delta_{\rm {\it C}\,F275W, F336W, F438W}$ and $\Delta_{\rm F275W, F814W}$ pseudo-colors, we ‘verticalized’ the two diagrams as in \cite{milone2017a}. We derived the ChM of NGC\,6838 in the same way but by using MS stars with  $17.75\le m_{\rm F814W} \le 18.75$.  

As illustrated in panels a3 and b3 of Figure~\ref{fig:plot_cmd},  the distribution of MS stars in the  $\Delta_{\rm {\it C}\,F275W, F336W, F438W}$ vs.\,$\Delta_{\rm F275W, F814W}$ plane is bimodal, in close analogy with what is observed for the RGB, indicating the presence of multiple populations among the MS.

\section{The extended first-generations of NGC 6362 and NGC 6838}
\label{sec:1g_extended}
To further investigate multiple populations of NGC\,6362 and NGC\,6838, we reproduce the ChMs of RGB stars and the $\Delta_{\rm F275W,F814W}$ kernel-density distributions of 1G, 2G, and observational errors in the upper panels of Figure~\ref{fig:ChMs}. Stars near the origin of the ChM reference frame may reflect the chemical composition of their natal cloud and correspond to the 1G, whereas the 2G component, which is He- and N-enhanced and C- and O-depleted, defines the sequence elongated towards large $\Delta_{\rm {\it C}\, F275W, F336W, F438W}$ and low $\Delta_{\rm F275W, F814W}$ values \citep[e.g.,][]{milone2020b, jang2021a}. Hence, the aqua dash-dotted lines, drawn by hand, separate 1G and 2G stars that we colored red and blue, respectively. Both 1G and 2G stars of each cluster span wider $\Delta_{\rm F275W,F814W}$ intervals than what is expected from observational errors alone, indicated by orange dots in Figure~\ref{fig:ChMs}. 
 
Moreover, 1G and 2G stars exhibit different patterns in the ChM. The kernel-density distributions of $\Delta_{\rm F275W,F814W}$ reveal that 2G stars in both clusters have nearly symmetric patterns and are centered around $\Delta_{\rm F275W,F814W}\sim-0.18$ mag. Conversely, the kernel-density distributions of 1G stars present more-complex behaviours with hints of multiple peaks. In  both clusters, for example, the frequencies of 1G stars suddenly increase from  $\Delta_{\rm F275W,F814W} \sim -0.20$ to $-0.15$ mag, where they approach their maximum values, and gently decline towards $\Delta_{\rm F275W,F814W} \sim 0.0$ mag. Some hints of secondary  peak(s) are visible towards $\Delta_{\rm F275W,F814W} \sim -0.1$ -- $0.0$. Clearly, in both clusters 1G stars display more extended ChM sequences than the 2G. 

In the following, we investigate whether the extended 1G sequences are peculiarities of the RGB or are also present among unevolved MS stars. To do this, we exploit multi-band {\it HST} photometry to analyze the color distributions of 1G stars along the MSs of NGC\,6362 and NGC\,6838. In the next subsection we demonstrate that MS 1G stars of both GCs exhibit intrinsic color extensions (subsection~\ref{subsec:ChM_1g}). We then continue our analysis of the 1G color spread exploring the physical phenomenon responsible for it. Specifically, in subsection~\ref{subsec:binaries} we consider the role of binaries in shaping the 1G ChM sequence, whereas in subsection~\ref{subsec:1g_synth} we constrain the chemical composition of the 1G by considering MS stars. Finally, in subsection~\ref{subsec:distribuzioni} we use the $\Delta_{\rm F275W,F814W}$ pseudo-color broadening to derive the metallicity distribution of 1G stars. 

\subsection{Chemical inhomogeneities among 1G MS stars}
\label{subsec:ChM_1g}
Lower panels of Figure~\ref{fig:ChMs} reproduce the ChMs for MS stars in NGC\,6362 and NGC\,6838 together with the $\Delta_{\rm F275W,F814W}$ kernel-density distributions of 1G, 2G, and observational errors. Though the 1G-2G separation in MS ChMs is not as sharp as in the RGB, due to higher photometric errors tending to fill the gap, we were still able to identify two distinct sequences on the MS ChM of both clusters, corresponding to the 1G and 2G component. The aqua dash-dotted line, drawn by hand to separate the two main stellar populations, defines the samples of bona-fide 1G and 2G stars, marked with red and blue colors, respectively. 

\begin{figure*} 
  \centering
  \includegraphics[height=10cm,trim={0.3cm 2.9cm 1.6cm 6.6cm},clip]{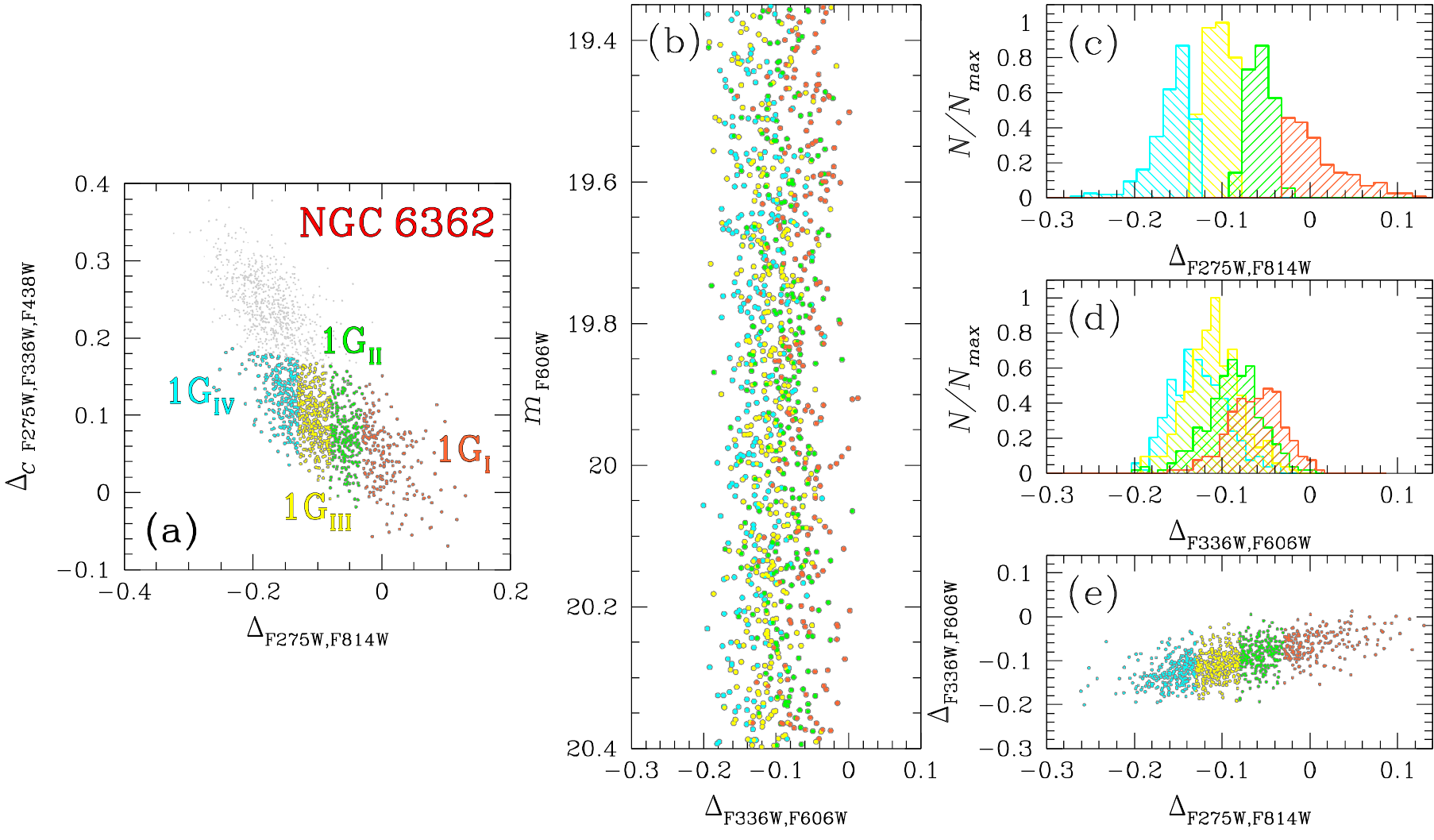}
  \caption{\emph{Panel a.} ChM of MS stars in NGC\,6362 where we have highlighted in orange, green, yellow, and cyan the samples of 1G$_{\rm I--IV}$ stars, respectively. \emph{Panel b.} $m_{\rm F606W}$ vs.\,$\Delta_{\rm F336W,F606W}$ diagram where we have plotted only 1G stars identified on the ChM. \emph{Panel c and d.} Normalized histograms of the $\Delta_{\rm F275W,F814W}$ and $\Delta_{\rm F336W,F606W}$ distributions for each sample of 1G stars identified on the ChM. \emph{Panel e.} Correlation between the $\Delta_{\rm F336W,F606W}$ and $\Delta_{\rm F275W,F814W}$ pseudo-colors.}  
  \label{fig:plot_CMDs_1g}
\end{figure*}

Consistently with what is observed on the RGB ChMs, the 1G MS stars of NGC\,6362 and NGC\,6838 display a well-elongated distribution along the $\Delta_{\rm F275W, F814W}$ direction. Conversely, the 2G sequence exhibits a narrower $\Delta_{\rm F275W, F814W}$ spread on the MS ChM of both clusters, as highlighted by the corresponding kernel-density distribution. To investigate whether the F275W$-$F814W color distribution of 1G stars is entirely due to observational errors or it is intrinsically broad we exploited three distinct approaches.
\begin{itemize}
  \item The first evidence that 1G MS stars are not consistent with a simple stellar population is provided by the fact that the observed $\Delta_{\rm F275W, F814W}$ spread of NGC\,6362 ($\sigma^{\rm obs, 1G}_{\rm F275W,F814W}$=0.062$\pm$0.003 mag) and NGC\,6838  ($\sigma^{\rm obs, 1G}_{\rm F275W,F814W}$=0.086$\pm$0.003 mag) are significantly wider than the color broadening inferred from the simulated ChM of a simple population ($\sigma^{\rm sim}_{\rm F275W,F814W}$=0.026 mag and 0.022 mag, respectively, orange points in the bottom-left corner of lower panels in Figure~\ref{fig:ChMs}). To account for the inclination of the 1G sequence in the ChM, we rotated the ChM as in \citet[][see their Figure\,2]{milone2017a} in such a way that the abscissa of the new reference frame is parallel to the 1G sequence. We repeated the comparison between the observed pseudo-color distribution and the AS tests as described above but in the rotated reference frame. We found that the observed pseudo-color spreads are much wider than the broadening associated with observational errors, thus confirming the conclusion of an intrinsic color spread in 1G stars.
  
  \item As highlighted by the comparison of the $\Delta_{\rm F275W, F814W}$ kernel distributions of 1G and 2G stars, the 1G exhibits wider color broadening than the 2G in both clusters ($\sigma^{\rm obs, 2G}_{\rm F275W,F814W}$=0.038$\pm$0.003 and 0.049$\pm$0.003 for NGC\,6362 and NGC\,6838, respectively). Since similar observational uncertainties affect the two stellar populations, this fact proves that the 1G is intrinsically broad.  
  
  \item To emphasize that the color spread of 1G MS stars observed on the ChM is intrinsic, we extended to NGC\,6362 and NGC\,6838 the procedure used by \cite{anderson2009} to prove that the MS of the GC 47\,Tucanae is composed of multiple populations. Their method is based on the idea that, if the MS broadening is due to observational errors alone, a star that is red (or blue) in a given photometric diagram has the same probability of being either red or blue in other diagrams built from a different dataset. On the contrary, separated color distributions are the signature of an intrinsic color spread. \\
 The procedure is illustrated in Figure~\ref{fig:plot_CMDs_1g} for NGC\,6362. We identified by eye in panel a four groups of 1G stars with different $\Delta_{\rm F275W,F814W}$ values, namely 1G$_{\rm I--IV}$, with the criteria that each group comprises about one fourth of 1G stars. The same stars are plotted in the other panels of Figure~\ref{fig:plot_CMDs_1g}. Specifically, panel b of Figure~\ref{fig:plot_CMDs_1g} shows the $m_{\rm F606W}$ vs.\,$\Delta_{\rm F336W,F606W}$ diagram, which is derived from a distinct dataset than that shown in panel a. Clearly, the four stellar groups, identified in panel a, have, on average, different $\Delta_{\rm F336W,F606W}$ pseudo-colors, thus confirming that the color spread is intrinsic. The fact that the average colors of the four stellar groups remain well separated is further illustrated in panels c and d, where we show the $\Delta_{\rm F275W,F814W}$ and $\Delta_{\rm F336W,F606W}$ histogram distributions, respectively. Finally, the correlation of the $\Delta_{\rm F336W,F606W}$  and $\Delta_{\rm F275W,F814W}$ pseudo-colors is provided in panel e of Figure~\ref{fig:plot_CMDs_1g}. The distribution of 1G stars in the $\Delta_{\rm F336W,F606W}$ vs.\,$\Delta_{\rm F275W,F814W}$  plane reinforces the conclusion that the 1G color spread is intrinsic. 
\end{itemize}

The discovery of extended 1G sequence among unevolved MS stars rules out the possibility that chemical inhomogeneities are the result of stellar evolution and suggests that 1G stars of GCs may be records of the chemical inhomogeneities in the original cloud where they formed at high redshift. Moreover, the fact that the 2G sequence displays a narrower color spread than the 1G on the ChM demonstrates that these inhomogeneities have been partially erased before the formation of 2G stars.

\begin{figure*} 
    \begin{center} 
    \includegraphics[height=15cm,trim={0cm 0.2cm .6cm 1cm},clip]{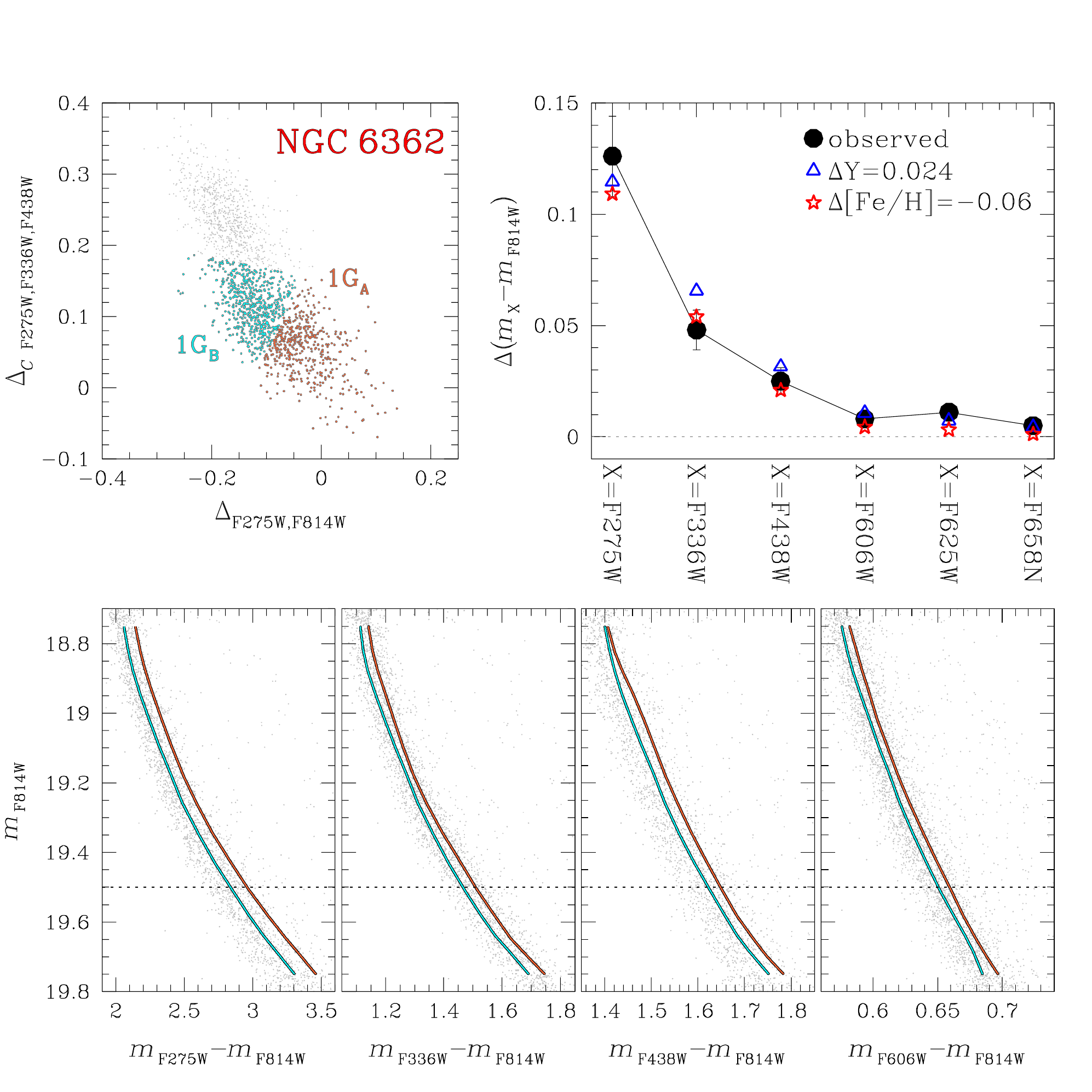}
    \caption{This figure illustrates the procedure used to derive the color differences between the two sub-groups of 1G$_{\rm A}$ and 1G$_{\rm B}$ stars for NGC\,6362. The upper-left panel reproduces the ChM of MS stars with the orange and cyan points that identify 1G$_{\rm A}$ and 1G$_{\rm B}$ stars, respectively. The lower panels show the $m_{\rm F814W}$ vs.\,$m_{\rm X}-m_{\rm F814W}$ CMDs where X=F275W, F336W, F438W, and F606W (the same has been done for F625W and F658N filters) in which the fiducial lines obtained for 1G$_{\rm A}$ and 1G$_{\rm B}$ are plotted in orange and cyan. In each panel, the horizontal dotted line corresponds to the reference F814W magnitude at which we calculate the color separation, $\Delta(m_{\rm X}-m_{\rm F814W})$. The upper-right panel shows the $\Delta(m_{\rm X}-m_{\rm F814W})$ calculated at $m^{\rm ref}_{\rm F814W}=19.5$ against the central wavelength of the X filter. The colors inferred from the best-fit synthetic spectra are overplotted with blue triangles and red starred symbols (see text for details).} \label{fig:chm_synth}
    \end{center}
\end{figure*} 
\subsection{The role of binaries in shaping the 1G color extension}
\label{subsec:binaries}
As illustrated in Figure~\ref{fig:ChMs}, the $\Delta_{\rm F275W,F814W}$ distributions of 1G stars along the RGB and the MS follow similar patterns. In particular, the distributions of both MS and RGB stars are not symmetric with respect to the peak, with a gentle decline towards $\Delta_{\rm F275W,F814W} \sim 0.0$. This feature would provide crucial constraints on the effect of binaries in shaping the ChM. Indeed, binary systems composed of two MS stars would populate the reddest part of the MS thus contributing to the red tail of the ChM. On the contrary, pairs of RGB stars exhibit bluer colors than single RGB stars and are responsible for the blue tail in the ChM \citep[][]{marino2019b}. 

The fact that the $\Delta_{\rm F275W,F814W}$ distributions of both MS and RGB exhibit a tail towards redder color demonstrates that unresolved binaries provide a minor contribution to the extension of the 1G sequence in the ChM. This finding is consistent with previous conclusion based on radial velocities of 1G stars \citep{marino2019b, kamann2020a} and supported by studies on the comparison of the observed 1G of NGC\,6752 with appropriate isochrones \citep{martins2020}. 

\subsection{Insights from multi-band photometry of MS stars}
\label{subsec:1g_synth}
To shed light on the physical phenomenon that is responsible for the color extension of the 1G sequence on the ChM, we applied to 1G stars of NGC\,6362 and NGC\,6838 the procedure by \cite{milone2012_47tuc, milone2018a}. This method, 
shown in Figure~\ref{fig:chm_synth} for NGC\,6362, consists in comparing the observed color widths of 1G stars with the predictions from appropriate isochrones that account for variations in light elements, helium, and iron. 

We first identified  two groups of 1G stars along the direction of the 1G color spread on the ChM, namely 1G$_{\rm A}$ and 1G$_{\rm B}$, with the criteria that each sample comprises about half of 1G members. 1G$_{\rm A}$ and 1G$_{\rm B}$ stars are respectively marked with orange and cyan dots in the upper-left panel of Figure~\ref{fig:chm_synth}. We then selected these two groups of 1G stars in the $m_{\rm F814W}$ vs.\,$m_{\rm X}-m_{\rm F814W}$ CMDs, where X=F275W, F336W, F438W, F606W, F625W, and F658N and derived the corresponding fiducial line. To do this, we divided the MS into a series of $m_{\rm F814W}$ bins of size $\delta m$ defined over a grid of points separated by fixed intervals of magnitude ($\rm \delta m/4$). For each bin the median $m_{\rm F814W}$ magnitude and  $m_{\rm X}-m_{\rm F814W}$ color have been calculated and then smoothed by means of boxcar averaging, where each point has been replaced by the average of the three adjacent points. The fiducial lines for 1G$_{\rm A}$ and 1G$_{\rm B}$ stars were derived by linearly interpolating the resulting points and are plotted with orange and cyan lines, respectively, in the lower panels of Figure~\ref{fig:chm_synth}.  

Subsequently, we defined a list of N reference points along the MS regularly spaced in $m_{\rm F814W}$ by 0.25. For each point, \emph{i}, we calculated the $\Delta (m_{\rm X}-m_{\rm F814W})$ color difference between the fiducial of 1G$_{\rm A}$ and 1G$_{\rm B}$ stars. The $\Delta (m_{\rm X}-m_{\rm F814W})$ values derived for $m_{\rm F814W, \emph{i}}=19.5$ (dotted horizontal line in bottom panels of Figure~\ref{fig:chm_synth}) in NGC\,6362 are represented as filled circles in the upper-right panel of Figure~\ref{fig:chm_synth}. Clearly, the color separation between 1G$_{\rm A}$ and 1G$_{\rm B}$ stars changes with the width of the color baseline reaching its maximum value for X=F275W. 

The $\Delta (m_{\rm X}-m_{\rm F814W})$ values measured for each filter were then compared with the colors inferred from a grid of synthetic spectra with appropriate chemical compositions. To derive them, as a preliminary step, we estimated gravity and effective temperature corresponding to each point, \emph{i}, by using the isochrones from the Dartmouth database \citep[][]{dotter2008} and adopting the same age, reddening, distance modulus, metallicity, and [$\alpha$/Fe] listed in \cite{dotter2010} for NGC\,6362 and NGC\,6838, respectively. For each reference point \emph{i}, we then used the codes ATLAS12 and SYNTHE \citep[][]{kurucz2005, castelli2005, sbordone2007} to compute a reference synthetic spectrum and a grid of comparison spectra with different chemical compositions. 

To calculate the reference one, we used the values of effective temperature, gravity, and metallicity (Z) derived previously from the best-fit isochrone, together with helium, $\rm Y=0.245+1.5\times Z$, solar abundances of C and N, and $\rm [O/Fe]=0.40$. Comparison spectra, instead, were derived by using the same chemical composition as the reference one but changing the abundances of He, C, N, O, and Fe. Specifically, in close analogy with \cite{milone2018a}, we simulated a grid of spectra enhanced in [N/Fe] up to 1.5 dex in steps of 0.1 dex, with [O/Fe] ranging from $-0.2$ to 0.4 in steps of 0.1 dex, [C/Fe] from $-0.4$ to 0.0 in steps of $-$0.1. Helium has been changed up to 0.33 and [Fe/H] by $\pm$0.2 dex. 

The atmospheric parameters, gravity, and effective temperature, of the reference spectra have been derived from the Dartmouth isochrones \citep[][]{dotter2008}, accordingly with their metallicity and helium content. The color differences $\Delta (m_{\rm X}-m_{\rm F814W})^{\rm synth}$ between the comparison and reference spectrum have been determined from the integration of each spectrum over the transmission curves of the seven {\it HST} filters used in this work for NGC\,6362. We applied the same procedure to NGC\,6838 with the only exception that for this cluster magnitudes in F625W and F658N filters were missing.

Results from the analysis of NGC\,6362 1G stars with $m_{\rm F814W, \emph{i}}=19.5$ are illustrated in the upper-right panel of Figure~\ref{fig:chm_synth}, where we compare the observed color differences between 1G$_{\rm A}$ and 1G$_{\rm B}$ stars with those inferred from the best-fit synthetic spectra. 
We found that colors derived from spectra with different light-element abundances provide poor fit to the observations. We conclude that 1G stars of NGC\,6362 share the same abundances of C, N, and O within our uncertainty of $\sim$0.1 dex.  As a consequence, for a fixed luminosity, they exhibit internal variations in their effective temperature. This fact indicates that the color broadening of the 1G component is due either to helium ($\rm \Delta Y=0.024$) or iron ($\rm \Delta[Fe/H]=0.06$ dex) variations. Indeed, the color differences inferred from spectra of 1G$_{\rm B}$ and 1G$_{\rm A}$ stars that differ either in [Fe/H] and or in Y alone provides a good match with the observed color differences (upper-right panel of Figure~\ref{fig:chm_synth}).  Similar results have been found for NGC\,6838, where the 1G is consistent with the presence of two groups of stars with either metallicity variations of $\rm \Delta [Fe/H]=0.05$ dex or helium differences of $\rm \Delta Y=0.020$\footnote{To demonstrate that the conclusions on internal chemical variations do not depend on colors that involve the F814W filter, we repeated the analysis based on $m_{\rm X}-m_{\rm F606W}$ (or $m_{\rm F606W}-m_{\rm X}$) colors. We verified that we obtain identical results on the internal variations in iron and helium deriving colors based on the F606W filter instead of the F814W one.}.

Our results, based on multi-band photometry of 1G MS stars, together with similar findings, based on  spectroscopy of 1G RGB stars \citep[e.g.,][]{marino2019a, marino2019b, kamann2020a}, indicate that 1G stars have constant light-element abundances.

\subsection{The metallicity distribution of 1G stars}
\label{subsec:distribuzioni}
In this subsection, we derive the [Fe/H] distribution of 1G stars in NGC\,6362 and NGC\,6838, by assuming that the $\Delta_{\rm F275W,F814W}$ pseudo-color broadening of the 1G is entirely due to metallicity variations. To do this, we adapted to the ChMs of RGB and MS stars, the method by \citet[]{stauffer1980a} and \citet{cignoni2010a} by comparing the observed $\Delta_{\rm F275W,F814W}$ distribution of 1G stars and the corresponding distributions of grids of simulated ChMs, where all stars have nearly-pristine helium content (Y=0.249) and the same light-element abundances. In a nutshell, we simulated the ChMs of N stellar populations with different metallicities in the interval between [Fe/H]$_{0}-0.5$ and [Fe/H]$_{0}+0.5$ dex, where [Fe/H]$_0$ is the iron abundance corresponding to $\Delta_{\rm F275W,F814W}$=0. We assumed [Fe/H]$_0$=$-1.1$ for NGC\,6362 and [Fe/H]$_0=-0.8$ for NGC\,6838, as inferred from high-resolution spectroscopy \citep[][]{massari2017a, carretta2009b}. We adopted for each stellar population a Gaussian [Fe/H] distribution with a tiny iron dispersion ($\sigma=0.01$ dex) and assumed that two adjacent populations differ in their average iron abundance by 0.01 dex.

We combined together the simulated populations and derived the $\Delta_{\rm F275W,F814W}$ histogram distribution in such a way that each population, j, contributes by a factor $c_{\rm j}$ to the total number of stars in the combined histogram. Here, $c_{\rm j}$  is a coefficient that ranges from 0 to 1. The simulated $\Delta_{\rm F275W,F814W}$ histogram distributions are then compared with the observed ones by means of Poissonian $\chi^{2}$ minimization:
\begin{equation}
     \chi^{2}=\sum_{\rm i}^{N_{\rm bins}} n_{\rm i} ln (n_{\rm i}/m_{\rm i}) - n_{\rm i} + m_{\rm i}
\end{equation}
where $N_{\rm bins}$ is the number of pseudo-color bins, while $n_{\rm i}$ and $m_{\rm i}$ are the numbers of observed and simulated 1G stars, respectively, in each bin. The  $\chi^{2}$ minimization has been performed with the \texttt{geneticalgorithm} Python public library\footnote{\url{https://pypi.org/project/geneticalgorithm/}} and the output consists in an array of coefficients $c_{\rm j}$ that provide the contribution of each population to the best-fit ChM of 1G stars. 
 
\begin{figure*} 
   \centering
   \includegraphics[height=8cm,trim={0.5cm 5cm 4.5cm 7.25cm},clip]{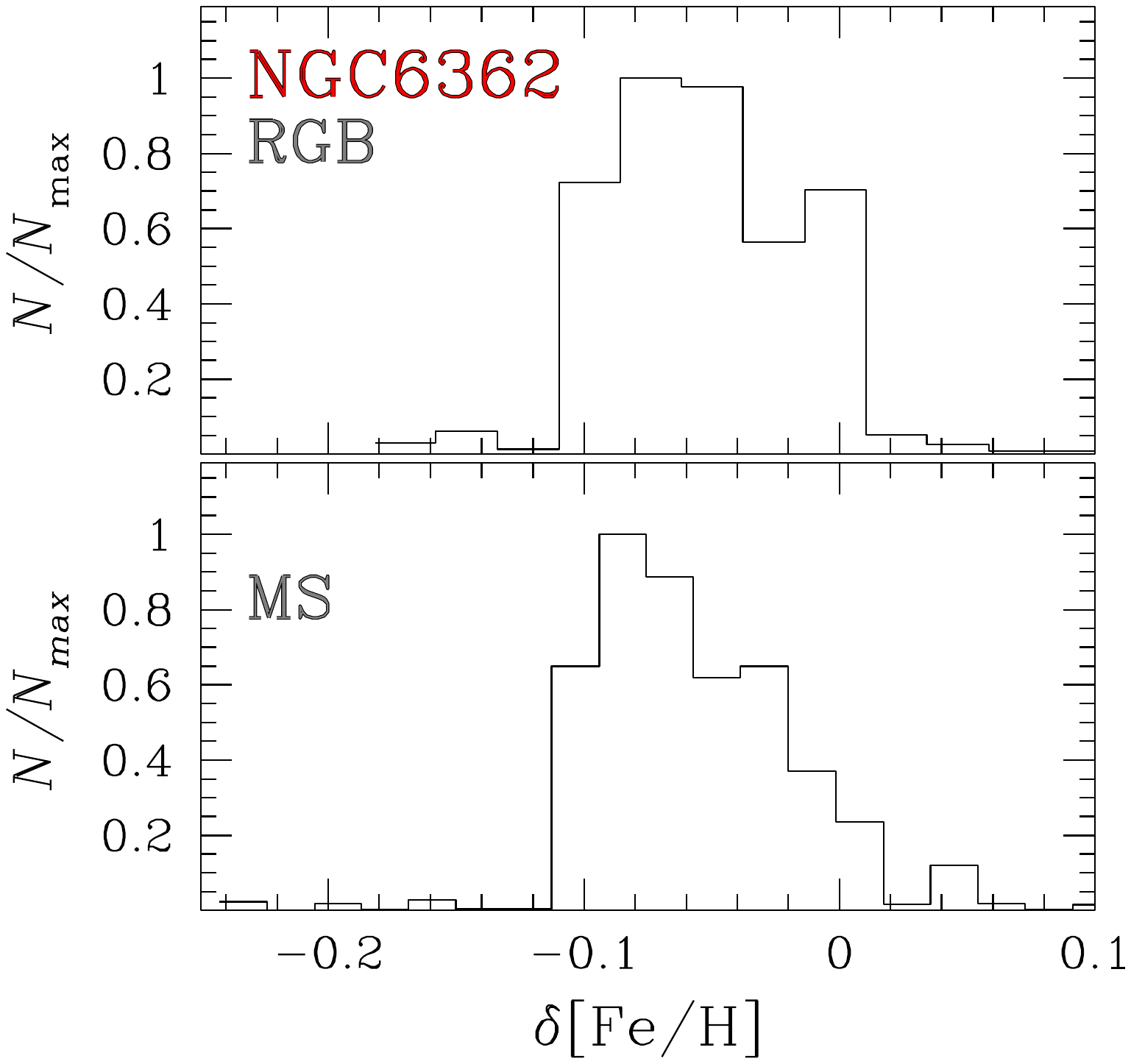}
   \includegraphics[height=7cm,trim={.5cm 8cm 7.5cm 9cm},clip]{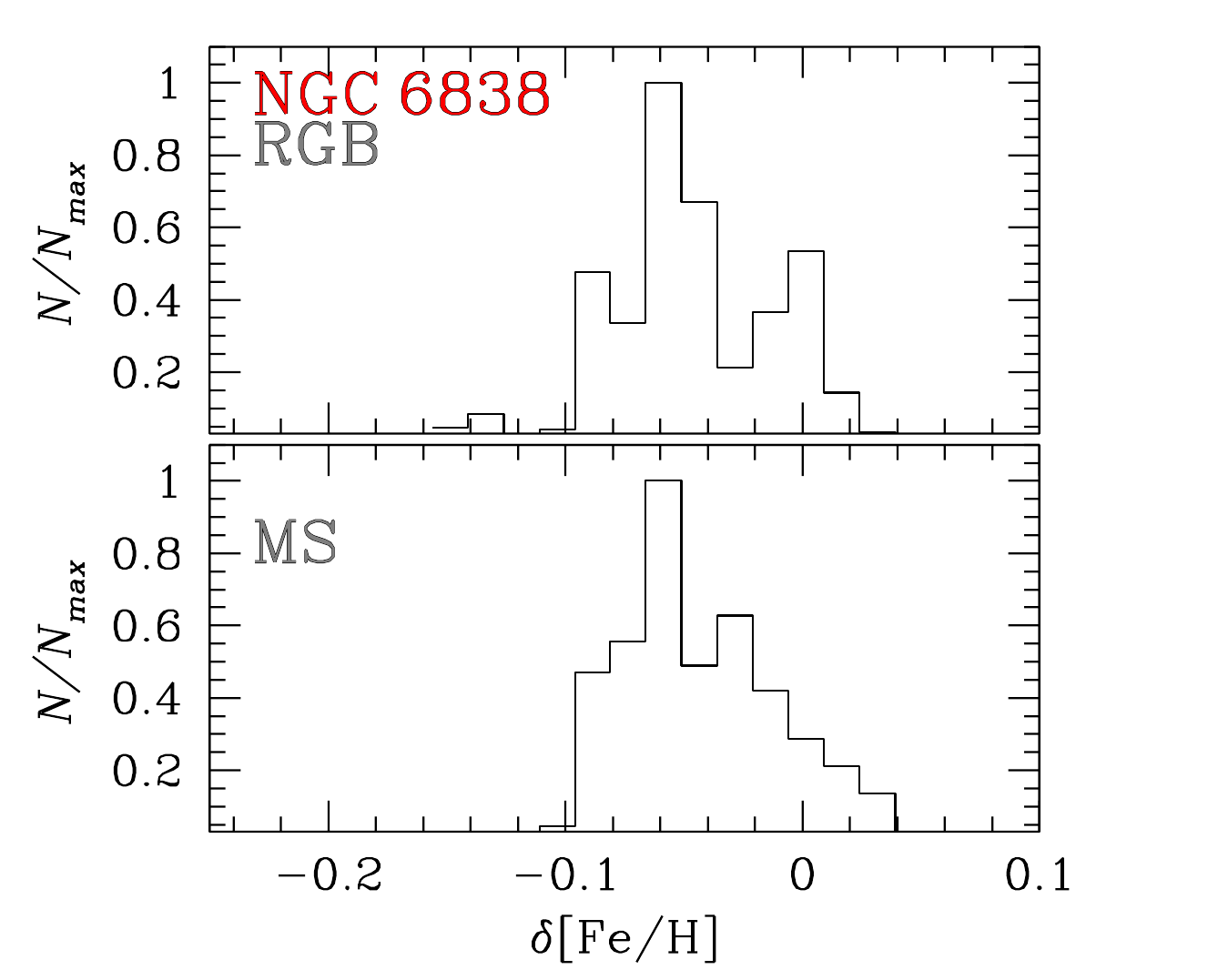}
   \caption{Histograms of the iron distribution of 1G stars in NGC\,6362 (left) and NGC\,6838 (right) inferred from MS stars (bottom) and RGB stars (top).} \label{fig:FeHist}
\end{figure*}

Results are illustrated in Figure \ref{fig:FeHist}, where we show the histogram distributions of the relative iron abundances of 1G stars, $\delta$[Fe/H], and show that we get comparable [Fe/H] distributions from RGB and MS stars that span similar intervals of 0.1 dex for NGC\,6362 and NGC\,6838.

\begin{figure*} 
    \begin{center} 
    \includegraphics[height=8cm,trim={0cm 11cm 0cm 0cm},clip]{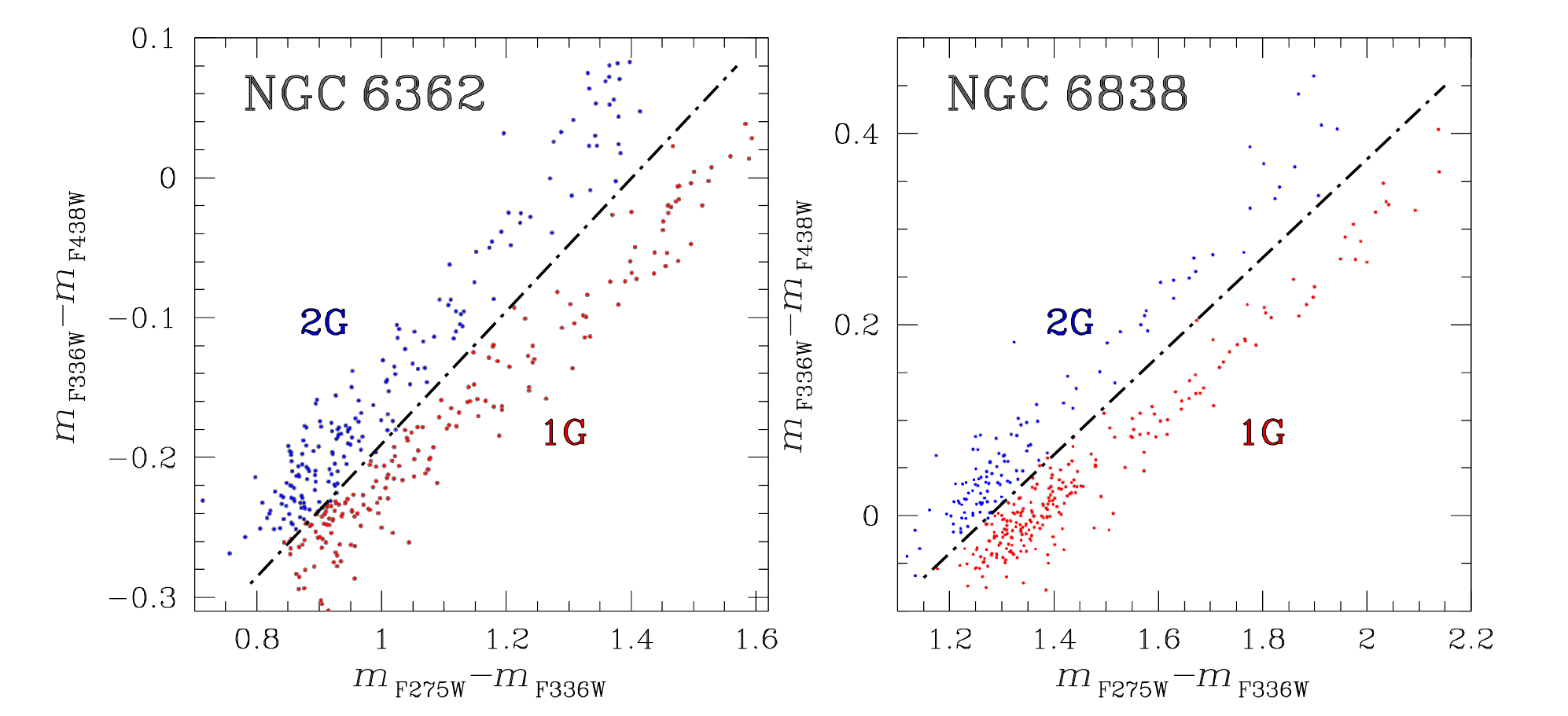}
    \caption{$m_{\rm F336W}-m_{\rm F438W}$ vs.\,$m_{\rm F275W}-m_{\rm F336W}$ two-color diagram for SGB stars of NGC\,6362 (left) and NGC\,6838 (right). The black dash-dotted line separates 1G and 2G stars, colored in red and blue, respectively} \label{fig:SGB_2color}
    \end{center}
\end{figure*} 
\begin{figure*} 
    \begin{center} 
    \includegraphics[height=11cm,trim={0.5cm 5cm 0.5cm 5.0cm},clip]{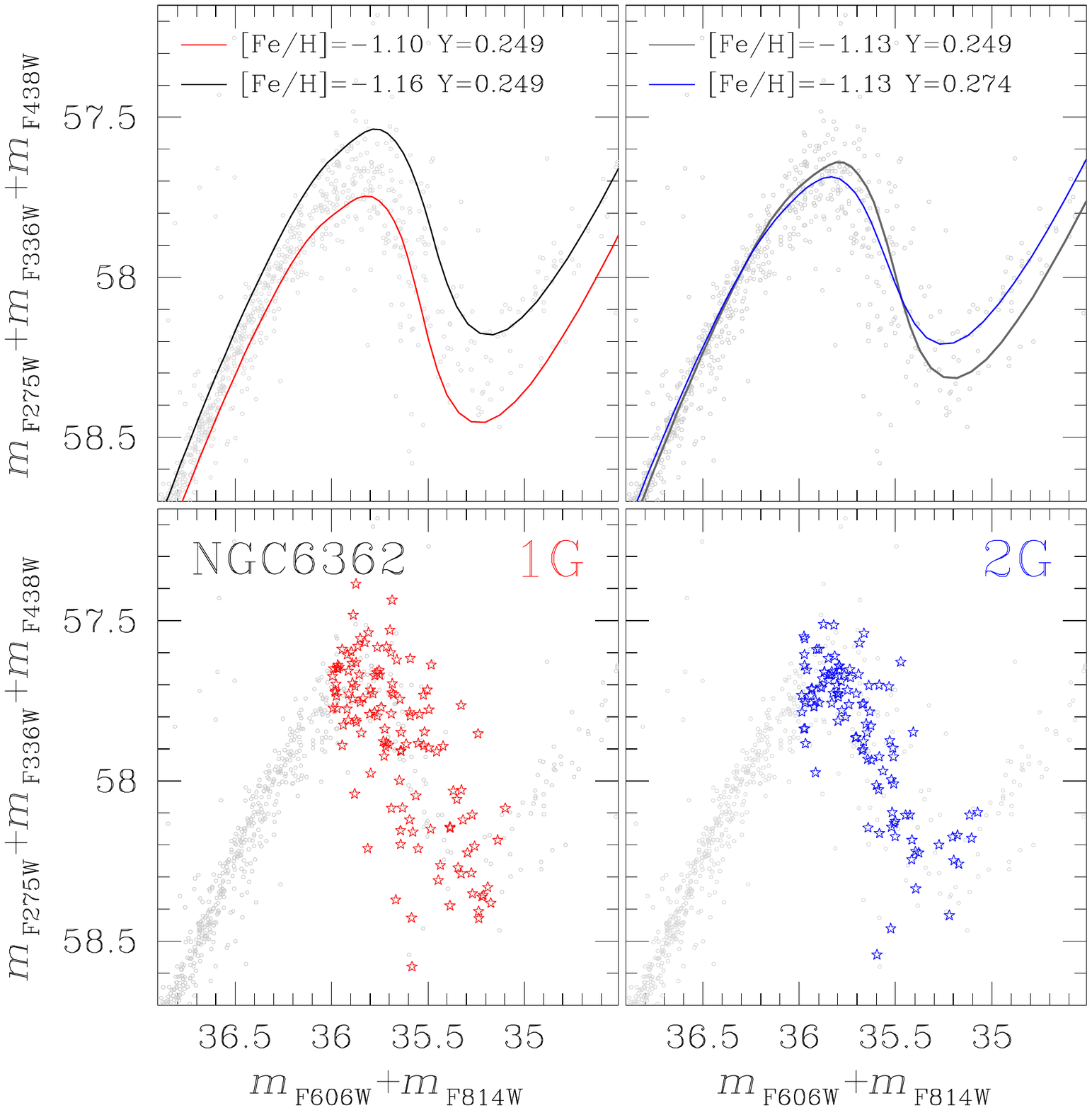}
    \caption{Pseudo two-magnitude diagrams for NGC\,6362 cluster members (grey dots) zoomed around the SGB. Upper panels compare isochrones from Dartmouth database \citep{dotter2008} with different helium content and metallicities. Lower panels mark 1G and 2G SGB stars selected in Figure \ref{fig:SGB_2color} with red and blue colors, respectively.} \label{fig:sgb6362}
    \end{center}
\end{figure*} 
\begin{figure*} 
    \begin{center} 
    \includegraphics[height=14cm,trim={0.5cm 5cm 2.5cm 4.0cm},clip]{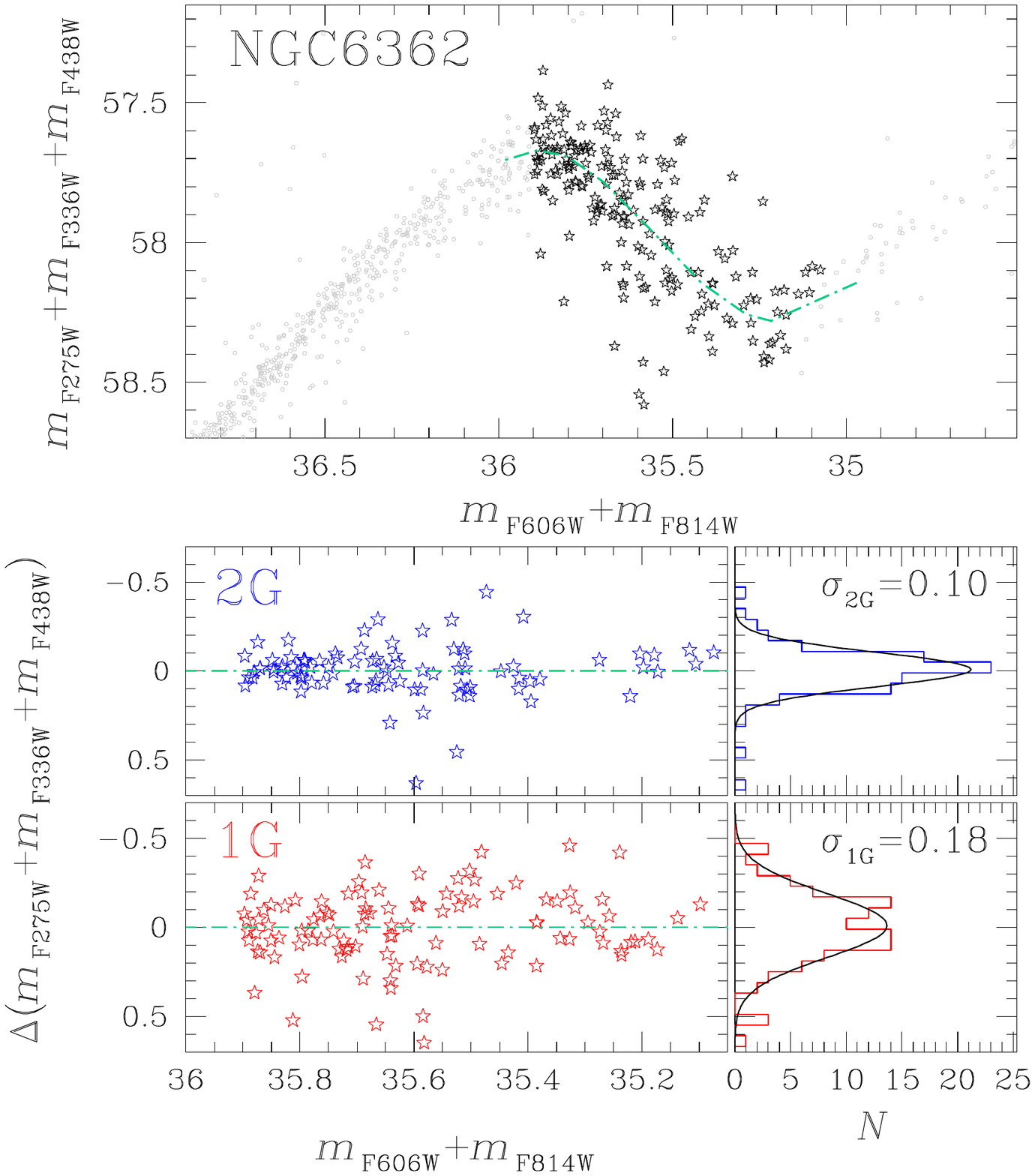}
    \caption{
    \textit{Upper panel.} Reproduction of the $m_{\rm F275W}+m_{\rm F336W}+m_{\rm F438W}$ vs.\,$m_{\rm F606W}+m_{\rm F814W}$ diagram of NGC\,6362 introduced in Figure~\ref{fig:sgb6362}. Black starred symbols mark the studied 1G and 2G SGB stars. The aqua dash-dotted line is the fiducial line of the SGB. 
    \textit{Lower panels.} Normalized $\Delta$($m_{\rm F275W}+m_{\rm F336W}+m_{\rm F438W}$) vs.\,$m_{\rm F606W}+m_{\rm F814W}$ diagram for 1G (bottom) and 2G stars (top). The corresponding $\Delta$($m_{\rm F275W}+m_{\rm F336W}+m_{\rm F438W}$) histogram distributions are plotted on the right together with the best-fit least-squares Gaussian functions.} \label{fig:NGC6362sgbV}
    \end{center}
\end{figure*} 

\section{Insights from the sub-giant branch}
\label{sec:1g_SGB}
To disentangle the effect of helium and metallicity variations on the color extension of the 1G component we introduce here a new approach based on SGB stars.

We first identified the sample of SGB cluster members in the $m_{\rm F814W}$ vs.\,$m_{\rm F438W}-m_{\rm F814W}$ CMD and plotted these stars in the $m_{\rm F336W}-m_{\rm F438W}$ vs.\,$m_{\rm F275W}-m_{\rm F336W}$ two-color diagram of Figure~\ref{fig:SGB_2color}. In this plane, SGB stars of both NGC\,6362 and NGC\,6838 exhibit a bimodal color distribution allowing us to identify the 1G and 2G components as the groups of stars on the right and on the left of the dash-dotted line, respectively. 

We focused on SGB stars because, in appropriate photometric diagrams, differences in helium and metallicity lead to different SGB morphologies, thus breaking the degeneracy observed in other evolutionary phases. As an example, the upper panels of Figure~\ref{fig:sgb6362} compare the photometry of NGC\,6362 stars with four alpha-enhanced isochrones ([$\alpha$/Fe]=0.4) from the  Dartmouth database \citep{dotter2008} in the $m_{\rm F275W}+m_{\rm F336W}+m_{\rm F438W}$ vs.\,$m_{\rm F606W}+m_{\rm F814W}$ plane. Specifically, the upper-left panel of Figure~\ref{fig:sgb6362} compares two isochrones with nearly-pristine helium abundance (Y=0.249) and different iron abundances ([Fe/H]=$-$1.10 and [Fe/H]=$-1.16$), whereas both isochrones plotted on the upper-right panel share the same metallicity [Fe/H]=$-$1.13 but different helium mass fractions, Y=0.249 and Y=0.274. The iron and helium differences match the values inferred from the color broadening of MS 1G stars. We adopted for all isochrones an age of 12.5 Gyr, distance modulus, ($m-M$)$_{0}$=13.34 mag, and reddening E(B$-$V)=0.07 mag, to match the observations of NGC\,6362. The SGBs of the two isochrones with different metallicities define parallel sequences, whereas the isochrones with different helium abundances exhibit smaller magnitude separations and cross to each other near the RGB base. Clearly, the isochrone behaviour justifies the choice of the adopted  pseudo magnitude-magnitude plane. Indeed, it maximizes the separation between isochrones with different iron content and allows us to disentangle the effect of helium and metallicity. 

Lower panels of Figure~\ref{fig:sgb6362} mark with red and blue starred symbols the two groups of 1G and 2G stars selected in Figure~\ref{fig:SGB_2color}, respectively. To quantify the magnitude broadening of SGB stars, we adopted the procedure illustrated in Figure~\ref{fig:NGC6362sgbV} for NGC\,6362. 

Briefly, we first derived the fiducial line of the SGB stars in the $m_{\rm F275W}+m_{\rm F336W}+m_{\rm F438W}$ vs.\,$m_{\rm F606W}+m_{\rm F814W}$ diagram plotted in the upper panel of Figure~\ref{fig:NGC6362sgbV}. To do this, we defined 0.15-mag wide intervals of $m_{\rm F606W}+m_{\rm F814W}$ and calculated the median values of $m_{\rm F606W}+m_{\rm F814W}$ and $m_{\rm F275W}+m_{\rm F336W}+m_{\rm F438W}$ in each bin. The fiducial is obtained by linearly interpolating these median points, and is represented with the aqua dash-dotted line. Finally, we calculated the $m_{\rm F275W}+m_{\rm F336W}+m_{\rm F438W}$ residuals of SGB stars, by subtracting to the $m_{\rm F275W}+m_{\rm F336W}+m_{\rm F438W}$ pseudo-magnitude of each star the corresponding value of the fiducial at the same $m_{\rm F606W}+m_{\rm F814W}$ value. The $\Delta$($m_{\rm F275W}+m_{\rm F336W}+m_{\rm F438W}$) values of 1G and 2G stars are plotted in the bottom-left panels of Figure~\ref{fig:NGC6362sgbV} as a function of $m_{\rm F606W}+m_{\rm F814W}$. The corresponding histogram distributions are shown in the bottom-right panels together with the best-fit least-squares Gaussian functions. Clearly, 1G SGB stars exhibit a wider pseudo-magnitude broadening than the 2G, as indicated by the standard deviation of the best-fit Gaussian functions of 0.18 and 0.10 mag, respectively. 

\begin{figure*} 
    \begin{center} 
    \includegraphics[height=11cm,trim={0.5cm 5cm 7.5cm 4.0cm},clip]{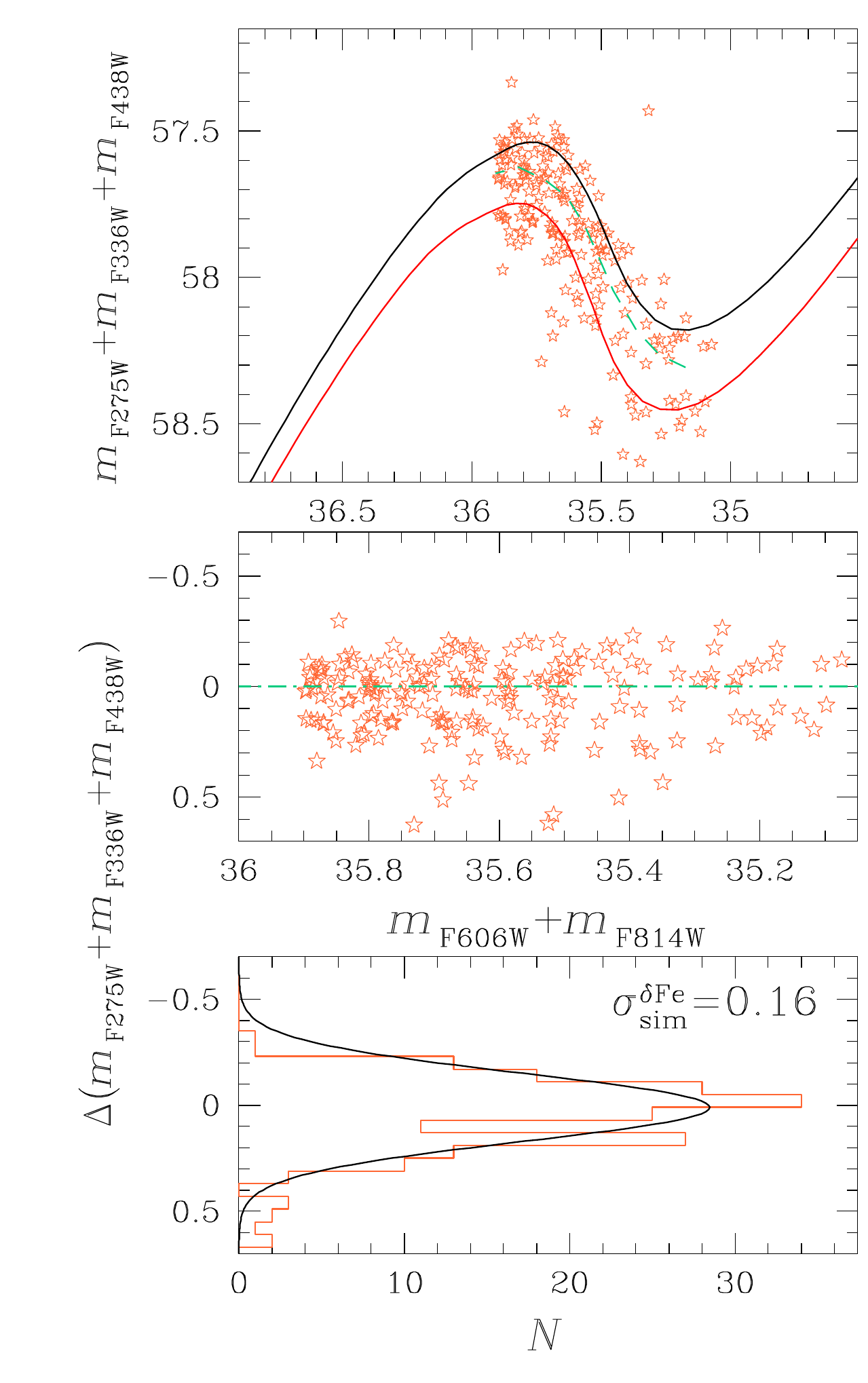}
    \includegraphics[height=11cm,trim={0.5cm 5cm 7.5cm 4.0cm},clip]{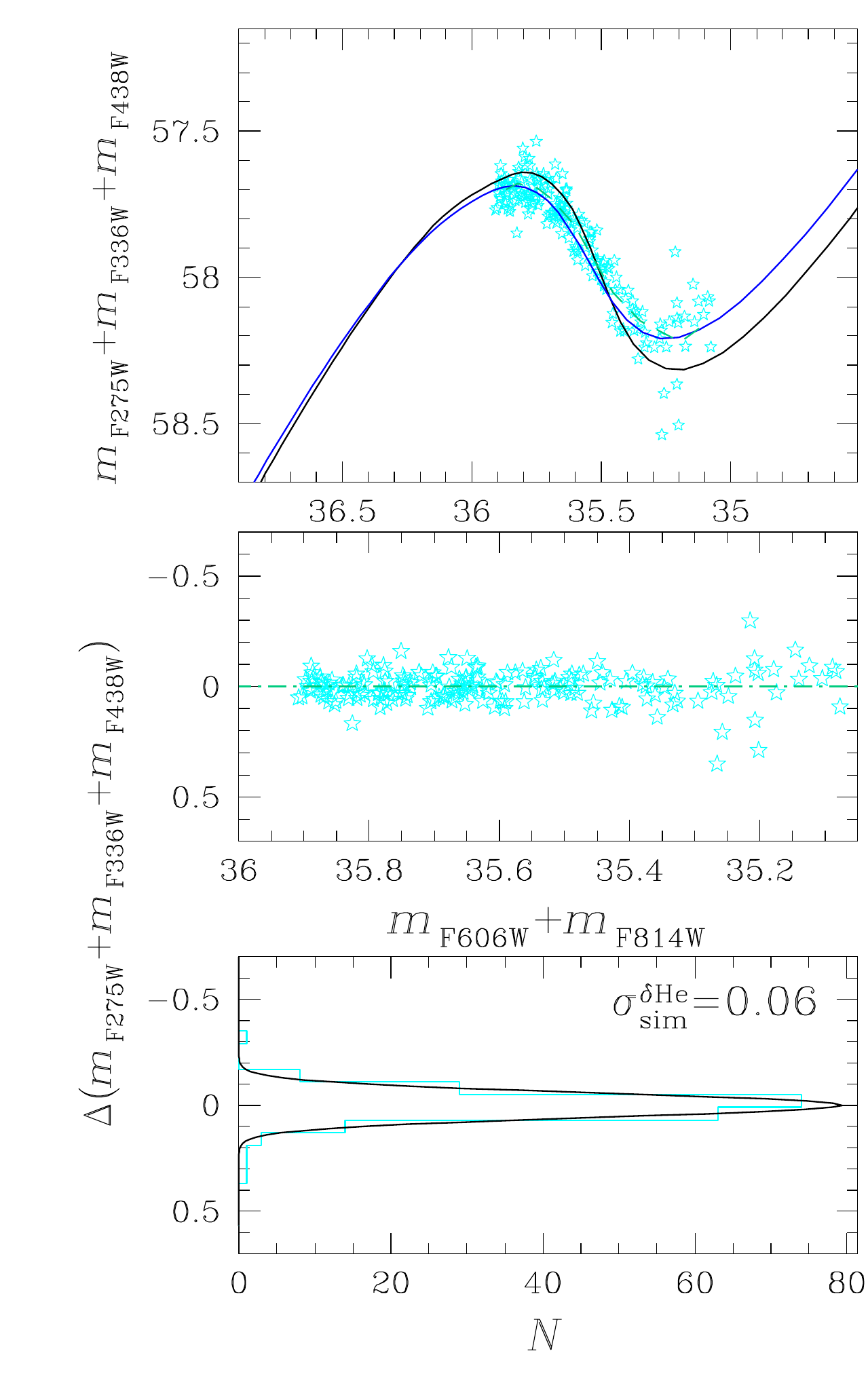}
    \caption{ \textit{Upper panels.} Simulated pseudo two-magnitude diagrams for SGB stars of composite stellar populations with the iron distribution (left panel) and helium distribution (right panel) inferred from 1G MS stars of NGC\,6362.
    The aqua dashed lines are the SGB fiducials.
    \textit{Middle panels.} $\Delta$($m_{\rm F275W}+m_{\rm F336W}+m_{\rm F438W}$) against $m_{\rm F606W}+m_{\rm F814W}$.
    \textit{Lower panels.} Histogram distributions of $\Delta$($m_{\rm F275W}+m_{\rm F336W}+m_{\rm F438W}$) for SGB stars. The best-fit least-squares Gaussian functions are represented with black lines and the corresponding $\sigma$ are quoted in each panel.} \label{fig:sgb6362SIMU}
    \end{center}
\end{figure*}

To investigate whether the magnitude broadening of 1G SGB stars is due to internal metallicity or helium variations\footnote{
An effect of changing the helium mass fraction (Y), is that for a fixed metal mass fraction (Z), a change in Y modifies the hydrogen mass fraction, thus affecting the metal-to-hydrogen ratio Z/X \cite[see][for details]{yong2013a}. As a consequence, increasing Y would result in rising [Fe/H]. However, it is unlikely that metallicity variations of NGC\,6362 and NGC\,6838 are induced by helium variations. Indeed, helium enhanced stars should be located at low values of $\Delta_{\rm F275W,F814W}$ \citep[e.g.,][]{marino2019a, milone2020a}, whereas metal-rich stars populate the ChM region with large $\Delta_{\rm F275W,F814W}$ values.}, we compared observations with simulated pseudo two-magnitude diagrams, as illustrated in Figure~\ref{fig:sgb6362SIMU} for NGC\,6362. We first used ASs to generate the synthetic $m_{\rm F275W}+m_{\rm F336W}+m_{\rm F438W}$ vs.\,$m_{\rm F606W}+m_{\rm F814W}$ diagram for stellar populations with Y=0.249 and the metallicity distribution inferred in Section \ref{subsec:distribuzioni} (upper-left panel of Figure~\ref{fig:sgb6362SIMU}). We adopted the isochrones from \citet[]{dotter2008} and the values of distance, reddening, and age introduced before for NGC\,6362. This diagram is used to derive the $m_{\rm F275W}+m_{\rm F336W}+m_{\rm F438W}$ residuals of SGB stars that we plotted against $m_{\rm F606W}+m_{\rm F814W}$ in the middle-left panel of Figure~\ref{fig:sgb6362SIMU} and used to derive the $\Delta$($m_{\rm F275W}+m_{\rm F336W}+m_{\rm F438W}$) histogram distribution, shown in the bottom-left panel of Figure~\ref{fig:sgb6362SIMU}. We found that the $\Delta$($m_{\rm F275W}+m_{\rm F336W}+m_{\rm F438W}$) distribution of simulated SGB stars is reproduced by a Gaussian function with dispersion $\sigma^{\rm \delta Fe}_{\rm sim}=0.16$, which is comparable with the corresponding broadening of 1G SGB stars ($\sigma_{\rm 1G}=0.18$). Similarly, we simulated the diagram shown in the upper-right panel of Figure~\ref{fig:sgb6362SIMU} adopting constant [Fe/H]=$-1.13$ and the helium distribution derived with the procedure described in Section~\ref{subsec:distribuzioni} but by assuming that the broadening of 1G stars is entirely due to helium variations. In this case, we obtained a much smaller value for the $\Delta$($m_{\rm F275W}+m_{\rm F336W}+m_{\rm F438W}$) dispersion of $\sigma^{\rm \delta He}_{\rm sim}=0.06$, which is not consistent with the observations of 1G stars. 

\begin{figure*} 
    \begin{center} 
    \includegraphics[height=14cm,trim={0.01cm 3cm 4.5cm 1.0cm},clip]{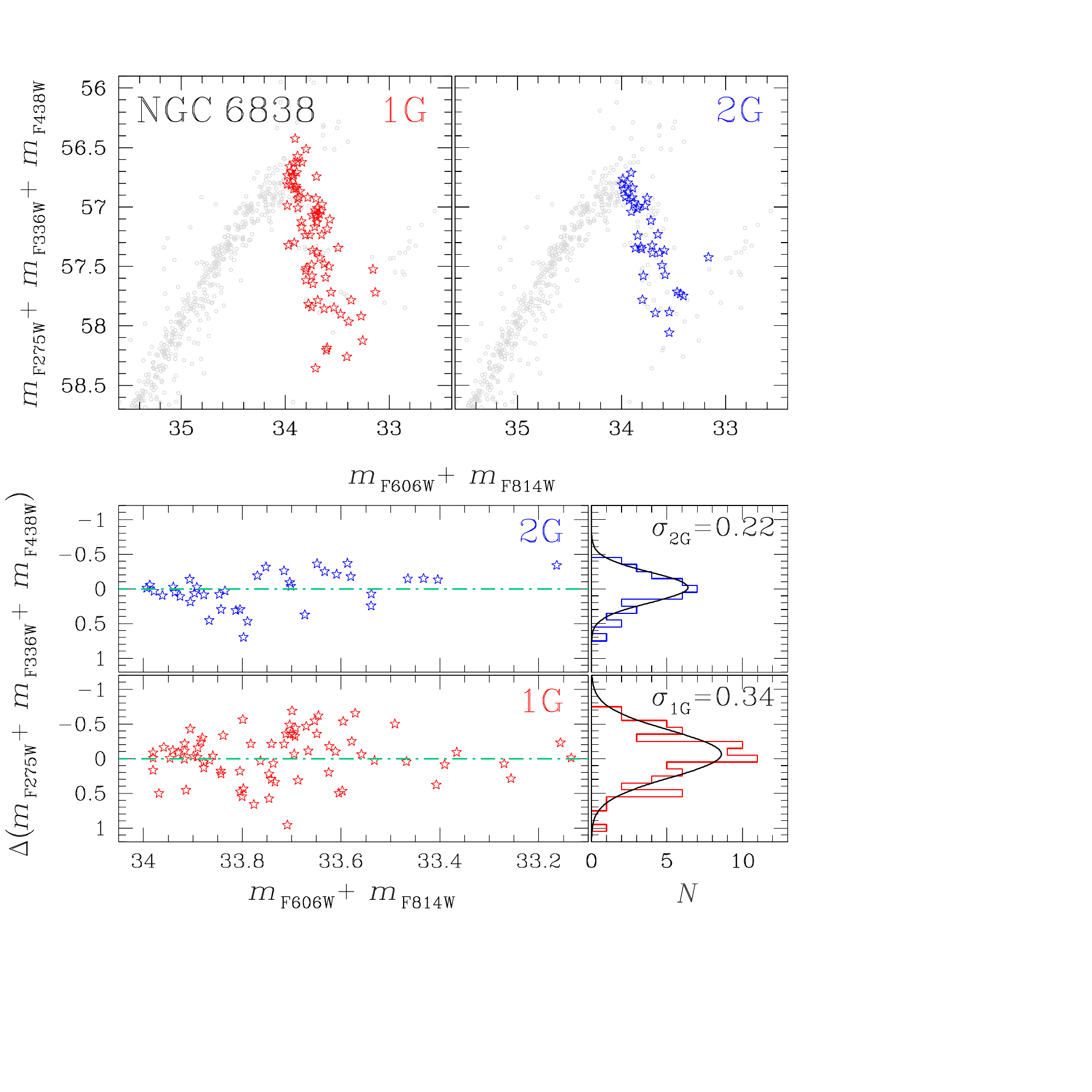}
    \caption{\textit{Upper panels.} Pseudo two-magnitude diagrams for NGC\,6838 cluster members (grey dots) zoomed around the SGB. 1G (left) and 2G (right) SGB stars selected in Figure~\ref{fig:SGB_2color} are marked with red and blue starred symbols, respectively. \textit{Lower panels.} Normalized $\Delta$($m_{\rm F275W}+m_{\rm F336W}+m_{\rm F438W}$) vs.\,$m_{\rm F606W}+m_{\rm F814W}$ diagram for 1G (bottom) and 2G stars (top). The corresponding $\Delta$($m_{\rm F275W}+m_{\rm F336W}+m_{\rm F438W}$) histogram distributions are plotted on the right together with the best-fit least-squares Gaussian functions.}  \label{fig:NGC6838sgbV}
    \end{center}
\end{figure*}

\begin{figure*} 
    \begin{center} 
    \includegraphics[height=7.5cm,trim={0.15cm 8cm 6.cm 11cm},clip]{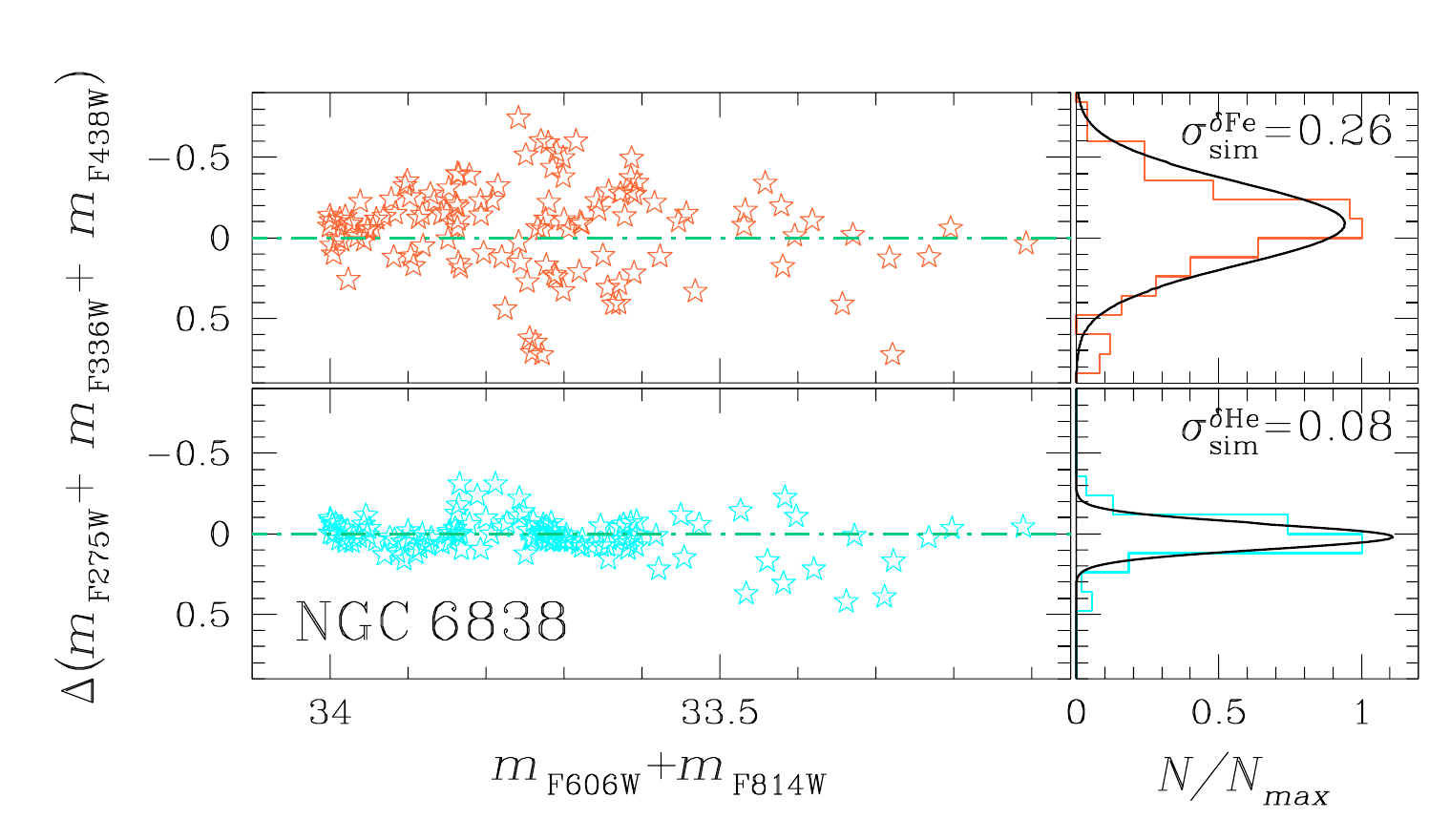}
    \caption{As in middle and lower panels of Figure~\ref{fig:sgb6362SIMU} but for NGC\,6838.}  \label{fig:NGC6838simu}
    \end{center}
\end{figure*}

We carried out the same analysis for NGC\,6838 SGB stars and summarized the results in Figure~\ref{fig:NGC6838sgbV}. The upper-panels represent the $m_{\rm F275W}+m_{\rm F336W}+m_{\rm F438W}$ vs.\,$m_{\rm F606W}+m_{\rm F814W}$ diagram where we marked 1G and 2G stars with red and blue starred symbols, respectively. The $\Delta(m_{\rm F275W}+m_{\rm F336W}+m_{\rm F438W})$ values of 1G and 2G stars are plotted as a function of $m_{\rm F606W}+m_{\rm F814W}$ in the bottom-left panels while the corresponding histogram distributions are shown in the bottom-right ones together with the best-fit least-squares Gaussian functions. In close analogy with NGC\,6362, 1G SGB stars of NGC\,6838 are characterized by a wider pseudo-magnitude broadening ($\sigma_{\rm 1G}=0.34$) than the 2G ($\sigma_{\rm 2G}=0.22$). Moreover, we generated a synthetic $m_{\rm F275W}+m_{\rm F336W}+m_{\rm F438W}$ vs.\,$m_{\rm F606W}+m_{\rm F814W}$ diagram by using the metallicity distribution derived from MS stars of NGC\,6838 in Section~\ref{subsec:distribuzioni}. This diagram has been then exploited to infer the $m_{\rm F275W}+m_{\rm F336W}+m_{\rm F438W}$ residuals of SGB stars and the correspondent histogram distribution, illustrated in the upper-left and -right panel of Figure~\ref{fig:NGC6838simu}, respectively. In this case, the $\Delta$($m_{\rm F275W}+m_{\rm F336W}+m_{\rm F438W}$) distribution is reproduced by a Gaussian function with dispersion $\sigma^{\rm \delta Fe}_{\rm sim}=0.26$, which is comparable with the corresponding broadening of 1G SGB stars. Conversely, exploiting the helium distribution to generate the synthetic $m_{\rm F275W}+m_{\rm F336W}+m_{\rm F438W}$ vs.\,$m_{\rm F606W}+m_{\rm F814W}$ diagram, the best-fit Gaussian function has a smaller dispersion ($\sigma^{\rm 
\delta He}_{sim}=0.08$) which is not consistent with observations of 1G stars, as illustrated in bottom panels of Figure~\ref{fig:NGC6838simu}. 

The results obtained from the analysis of SGB stars of NGC\,6362 and NGC\,6838 allow us to disentangle the effect of helium and metallicity and conclude that star-to-star metallicity variations are primarily responsible for the extended 1G sequence in the ChM. 


\section{Metallicity variations in Galactic Globular Clusters}
\label{sec:correlations}
In the previous Section we have demonstrated that the color extension of 1G stars observed on the ChMs of NGC\,6362 and NGC\,6838 is the signature of intrinsic metallicity variations. By extending this conclusion to all GCs, in the following we use the colors of 1G RGB stars to infer the maximum iron abundance spread within the 1G of 55 Galactic GCs. Moreover, we investigate the relations between the iron spread and the main parameters of the host GC. 
\begin{figure*} 
    \begin{center} 
    \includegraphics[height=5cm,trim={.7cm 10.5cm 1.cm 11.8cm},clip]{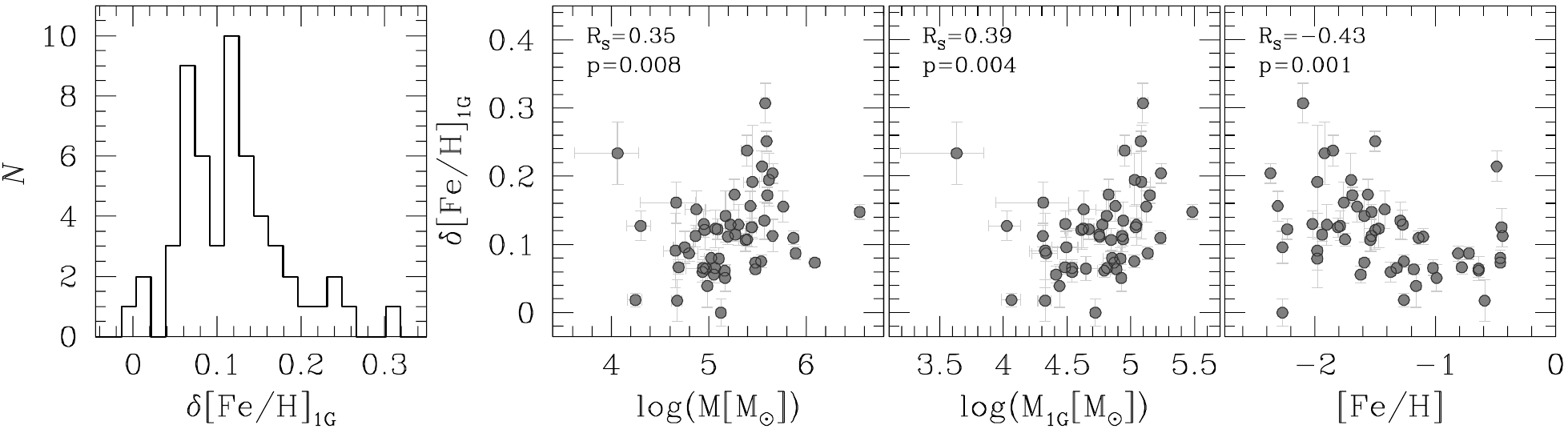}
    \caption{Histogram distribution of $\rm \delta [Fe/H]_{\rm 1G}$ (left) for 55 Galactic GCs. The relations between $\rm \delta [Fe/H]_{\rm 1G}$ and cluster mass, 1G mass, and [Fe/H] are plotted in the other three panels. The Spearman's rank correlation coefficient and the associated p-value are quoted in the upper-left corners.} \label{fig:hist_corrFe}
    \end{center}
\end{figure*} 

To quantify the $\rm \Delta_{F275W, F814W}$ extension of 1G stars observed on the ChM, \cite{milone2017a} have estimated the intrinsic color width of 1G RGB stars belonging to 55 Galactic GCs ($W^{\rm 1G}_{\rm F275W,F814W}$). By assuming that iron variations are the only responsible for the 1G color spread, we exploited the color-metallicity relations by \citet[]{dotter2008} to transform the values of $W^{\rm 1G}_{\rm F275W,F814W}$ into internal [Fe/H] variation ($\rm \delta [Fe/H]_{\rm 1G}$).

Results are listed in Table~\ref{Fe_table}. As illustrated in the left panel of Figure~\ref{fig:hist_corrFe}, where we show the $\rm \delta [Fe/H]_{\rm 1G}$ histogram distribution, the internal iron variation within the 1G component changes dramatically from one cluster to another ranging from $\sim 0.00$ to $\sim 0.30$, with an average value of $\rm \delta [Fe/H]_{\rm 1G} \sim 0.12$. 

\subsection{Relations with the host GC parameters}
In the following, we investigate the relations between the internal iron variations among 1G stars ($\rm \delta [Fe/H]_{\rm 1G}$) and the main parameters of the host GC. 

Overall, our analysis includes 20 global GC parameters. Iron abundance ([Fe/H]), absolute visual magnitude ($M_{V}$), ellipticity ($\epsilon$), central concentration (\emph{c}), central stellar density ($\rho_{0}$), and central surface brightness ($\mu_{\rm V}$) have been taken from the 2010 version of the Harris catalog \citep[]{harris1996}. Present-day ($\rm M$) and initial ($\rm M_{\rm in}$) mass of the host cluster, central escape velocity ($\rm v_{\rm esc}$), central velocity dispersion ($\sigma_{\rm v}$), core density ($\rho_{c}$), and core relaxation time at half mass ($\rm \tau_{hm}$) have been derived by \cite{baumgardt2018}. To compute the 1G mass (M$_{\rm 1G}$) we exploited the fraction of 1G stars measured in previous papers by \cite{milone2017a, milone2020b} and \cite{dondoglio2021a}. Finally, GC ages have been derived by \citet[MF09]{marinfranch2009}, \citet[D10]{dotter2010}, \citet[V13]{vandenberg2013}, and \citet[T20]{tailo2020}, while the fractions of binary stars in GCs have been calculated by \cite{milone2012redd} within the cluster core ($f^{\rm C}_{\rm bin}$), in the region between the core and the half-mass radius ($f^{\rm C-HM}_{\rm bin}$), and beyond the half-mass radius ($f^{\rm oHM}_{\rm bin}$).

For each pair of analyzed quantities we estimated the statistical correlation between the two by calculating the Spearman's rank coefficient (R$_{\rm S}$) and the corresponding p-value. The results of the correlations between $\rm \delta [Fe/H]_{\rm 1G}$ and the cluster parameters are listed in Table~\ref{correlations_table}, where for each couple of variables we provide R$_{\rm S}$, the corresponding p-value, and the number of degrees of freedom. 

As shown in Figure~\ref{fig:hist_corrFe}, we find mild correlations between  $\rm \delta [Fe/H]_{\rm 1G}$ and cluster mass and 1G mass (R$_{\rm S}=0.35$, and 0.39, respectively), with massive GCs having on average larger 1G metallicity spreads. Moreover, the internal iron variation of 1G stars decreases with cluster metallicity (R$_{\rm S}=-0.43$).
\begin{figure*} 
    \begin{center} 
     \includegraphics[height=10cm,trim={0.25cm 10cm 14.5cm 8.cm},clip]{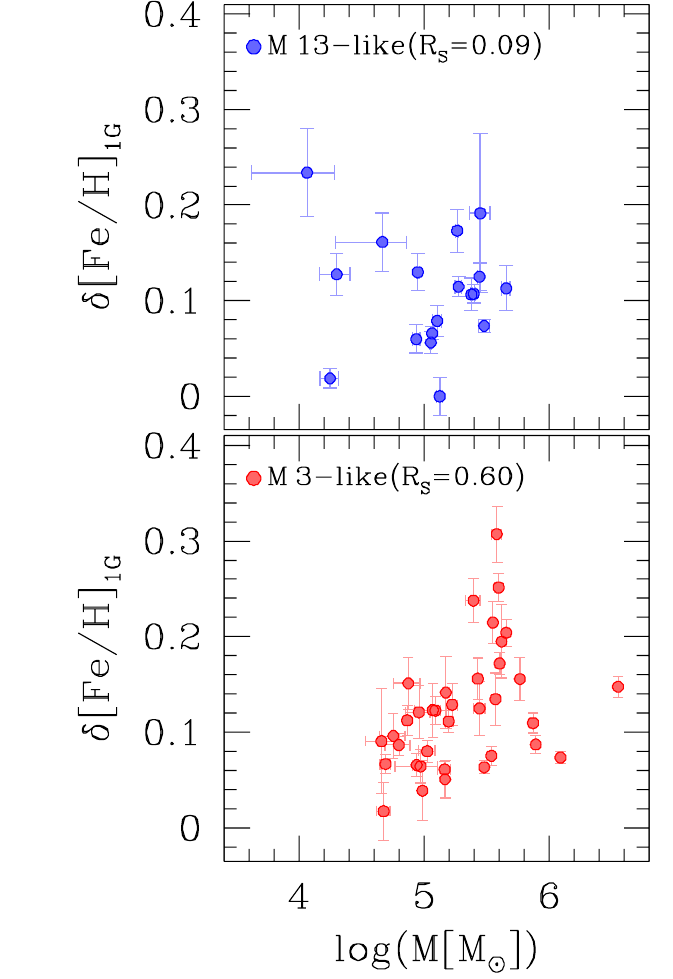}
    \includegraphics[height=10cm,trim={5cm 10cm 7.5cm 8cm},clip]{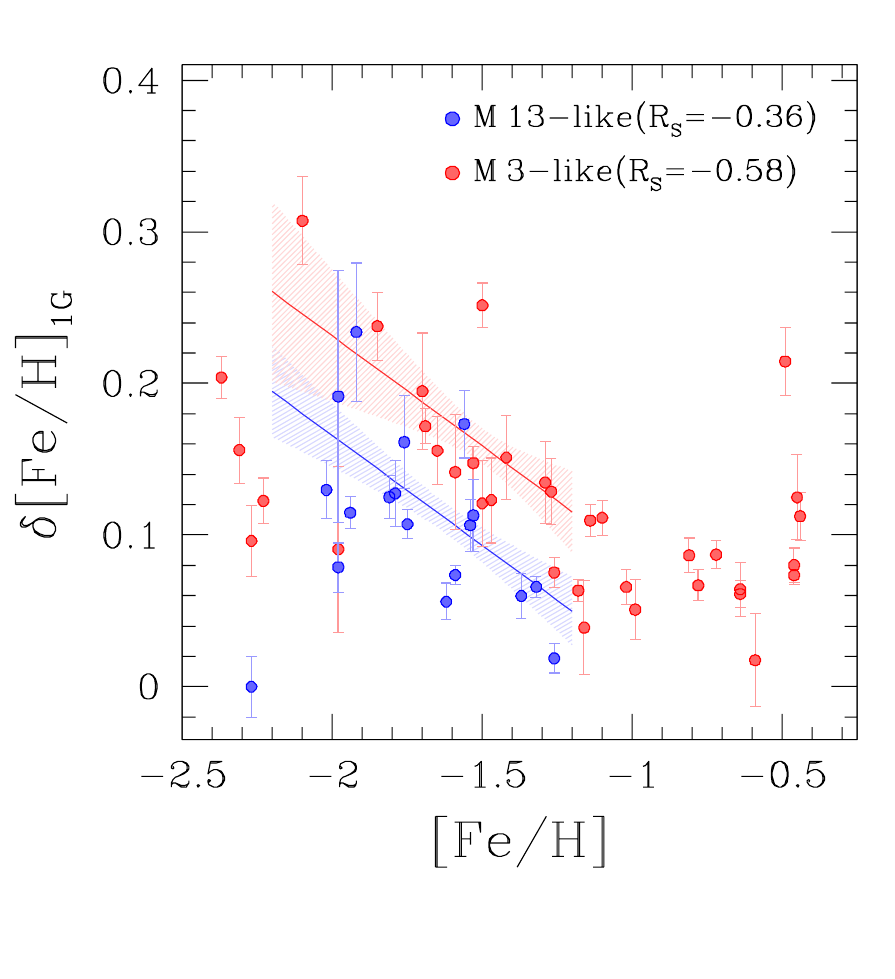}
    \caption{\textit{Left panels.} Internal iron variation of 1G stars, $\rm \delta [Fe/H]_{\rm 1G}$, against the logarithm of cluster mass for M\,13-like (top) and M\,3-like GCs (bottom). 
    \textit{Right panel.} $\rm \delta [Fe/H]_{\rm 1G}$ against [Fe/H]. Blue dots correspond to M\,13-like GCs while red ones to M\,3-like GCs. The best-fit least-squares straight lines derived for M\,13- and M\,3-like clusters with $-2.2<$[Fe/H]$<-1.2$ are colored blue and red, respectively. The Spearman's rank correlation coefficient is reported for each couple of quantities.} \label{fig:corrFe_M3M13}
    \end{center}
\end{figure*} 
\subsubsection{M3- and M13-like clusters}
To study further the relations between $\rm \delta [Fe/H]_{\rm 1G}$ and the host cluster parameters, we divided GCs in two categories of M\,3-like and M\,13-like GCs \citep[][]{milone2014, tailo2020}. This selection is based on the HB morphology and on the parameter L1, defined as the color difference between the RGB and the reddest part of the HB \citep[][]{milone2014}. Specifically, as in \cite{tailo2020}, we defined M\,3-like GCs with $\rm L_{1}\le0.35$ and M\,13-like GCs with $\rm L_{1}>0.35$. M\,3-like GCs mostly include GCs having red HBs while M\,13-like GCs present the blue HB only\footnote{Another distinctive feature of M\,13-like clusters is that their 1G stars suffer higher RGB mass loss than 1G stars of M\,3-like GCs and stars in simple-population GCs \citep{tailo2020, tailo2021a}. For this reason, the authors suggested that 1G stars of M\,13-like GCs form in high-density environments. As an alternative interpretation of the results from Tailo and collaborators, we can speculate that M\,13-like GCs have entirely lost their 1G stars as first suggested by \citet[][]{dantona2008a}. In this scenario, the stars that we name 1G are 2G stars that formed in a dense environment in the cluster center.}.

Intriguingly, the 1G iron variations of the two groups of GCs exhibit different behaviours. As illustrated in the left panels of Figure~\ref{fig:corrFe_M3M13},
there is no correlation between the $\rm \delta [Fe/H]_{\rm 1G}$ iron variation of M\,13-like GCs and cluster mass (R$_{\rm s}$=0.09), whereas M\,3-like GCs show a significant correlation with cluster mass (R$_{\rm s}$=0.60). Moreover, M\,13-like GCs display smaller iron variations than M\,3-like GCs with similar metallicity. To support this statement, we had least-squares fitted the $\rm \delta [Fe/H]_{\rm 1G}$ vs.\,[Fe/H] relation for M\,3- and M\,13-like clusters, respectively, with a straight-line, as shown in the right panel of Figure~\ref{fig:corrFe_M3M13}. To do that we considered only the common range of metallicity between the two groups of clusters, namely $-2.2<$[Fe/H]$<-1.2$, omitting GCs with [Fe/H]$\le-2.2$ that deviate from the common trend. The best-fit straight lines are given by $\rm \delta [Fe/H]_{\rm 1G}=(-0.139\pm0.056)\times[Fe/H]+(-0.058\pm0.093)$ (M\,3-like GCs, red line) and $\rm \delta [Fe/H]_{\rm 1G}=(-0.138\pm0.046)\times[Fe/H]+(-0.120\pm0.079)$ (M\,13-like GCs, blue line), respectively, thus corroborating the evidence that M\,13-like GCs exhibit smaller internal iron variations than M\,3-like GCs with similar metallicity. This fact is further confirmed by the most metal-poor clusters that we excluded from the fit. Indeed, the only M\,13-like object in this sample is characterized by the lowest $\rm \delta [Fe/H]_{\rm 1G}$ iron variation. 

\section{Summary and Discussion}
\label{sec:conclusion}
We have exploited {\it HST} multi-band images to present a photometric investigation of stellar populations within the Galactic GCs NGC\,6362 and NGC\,6838. The main purpose is to shed light on the physical reasons that are responsible for the extended 1G sequence in their ChMs.  Results can be summarized as follows:
\begin{itemize}
    \item We derived the ChM of MS stars of both GCs, where 1G and 2G stars define distinct sequences. We found that the 1G sequence has an intrinsic F275W$-$F814W color broadening in close analogy with what is observed for the RGB ChM. The evidence of extended 1G sequences among unevolved MS stars demonstrates that the color extension is not due to stellar evolution. Alternatively, this phenomenon could be the signature of chemical inhomogeneities in the proto-cluster cloud.

    \item We introduced the pseudo two-magnitude diagram $m_{\rm F275W}+m_{\rm F336W}+m_{\rm F438W}$ vs.\,$m_{\rm F606W}+m_{\rm F814W}$, where the SGBs of isochrones with different metallicities and helium abundances display different behaviours. The distribution of 1G SGB stars in these diagrams is consistent with internal variations in [Fe/H] but not with helium variations. The evidence that 1G SGB stars exhibit star-to-star iron variations, implies that the color extension of 1G sequences in the ChMs of 1G RGB and MS stars is due to metallicity variations up to $\delta$[Fe/H]$_{\rm 1G}=0.1$ dex.
\end{itemize}
Moreover, we compared the distributions of 1G and 2G stars along the ChM. Results have the potential to constrain the formation scenarios of multiple populations in GCs. 
\begin{itemize}

    \item We found that the F275W$-$F814W color extension of 2G MS stars of both NGC\,6362 and NGC\,6838 is significantly narrower than that of the 1G, but wider than the color broadening expected by observational errors. Specifically, subtracting in quadrature the effect of observational errors, the F275W$-$F814W spread of 2G stars is two times smaller than the corresponding 1G color spread. This fact demonstrates that the gaseous environment where 2G stars formed had more homogeneous iron content compared to the environment where 1G stars formed. 

    \item We exploited the 1G F275W$-$F814W color extension of the RGB from \citet[]{milone2017a} to estimate [Fe/H] variations within 1G stars of 55 GCs in the hypothesis that metallicity spread is the only responsible for the RGB broadening. The internal iron variation ranges from less than [Fe/H]$\sim$0.05 to $\sim$0.30 and mildly correlates with cluster mass and metallicity.

    \item Noticeably, the 2G sequences of NGC\,6362 and NGC\,6838 ChMs join the 1Gs on their metal-poor sides. In the hypothesis that the $\Delta_{\rm F275W,F814W}$ pseudo-color of 2G stars at the bottom of the 2G sequence is indicative of their iron abundance, these 2G stars would share the same metallicity as the metal-poor 1G stars. However, this is not an universal property of GCs. A visual inspection at the ChMs of Galactic GCs shows that the relative position between 1G and 2G sequences significantly changes from one cluster to another  \citep[see Figures 3--7 from][]{milone2017a}. While many clusters (e.g., NGC\,2808 and NGC\,6981) behave like NGC\,6362 and NGC\,6838, the 2G sequences of other clusters like NGC\,104 and NGC\,5272 are distributed on the metal-intermediate and, possibly, the metal-rich side of the 1G sequence.

    \item 1G stars of M\,13-like GCs, namely GCs with the blue HB only, present smaller iron variations than M\,3-like GCs with the same metallicity and no correlation with cluster mass. On the contrary, M\,3-like clusters exhibit significant correlation with cluster mass (R$_{\rm s}=0.60$).

\end{itemize}
The presence of metallicity variations among 1G stars might be due to the presence of metallicity variations within the ISM from which proto-cluster molecular clouds were born. Analytic models show that the metallicity scatter among the stars must however be lower than the scatter initially present in the gas \citep[e.g.,][]{Murray&Lin90, Bland-Hawthorn+10}. 
Turbulent diffusion within a proto-cluster cloud should smooth out chemical inhomogeneities at the scale of the cloud in roughly a crossing time -- defined as the ratio between the cloud radius and the cloud velocity dispersion -- and smaller-scale inhomogeneities even more quickly. Star formation in a molecular cloud also appears to begin in no more than a crossing time after its formation \citep[see][and references therein]{Krumholz+19}. This suggests that chemical inhomogeneities initially present in the gas should be reduced at the onset of star formation \citep[see also][]{Feng&Krumholz14, Armillotta+18}. In this picture, the presence of more or less metallicity variation in a given 1G population might be explained either by the presence of more or less metallicity variation in the original gas, or by different diffusion efficiencies during the cloud collapse. For example, we note that the cloud crossing time increases in low-density extended clouds, while decreases in high-density turbulent clouds.

Another possible explanation for the presence of metallicity variations among 1G stars is that the process of star formation within the cloud was still at play when the first supernovae exploded. In this picture, the cloud was nearly homogeneous at the onset of the star formation. Inhomogeneities arose at later times as supernovae polluted the surrounding star-forming gas. This possibility is explored in a recent work by \citet{mckenzie2021}. Through simulations of a gas-rich dwarf galaxy, the authors find that self-enrichment of star-forming clouds from short-lived massive stars may increase the metallicity dispersion within the resulting populations of 1G stars \citep[see also][]{Bailin18}. In their fiducial model, \citet{mckenzie2021} measure [Fe/H] abundance variations of the order of 0.1 dex, a value consistent with the average spread found in this paper. In comparison, He abundance spreads are found to be of the order of 0.001 dex, in agreement with the result that He alone is insufficient in elongating the 1G population in the ChM. However, as pointed out by \citet{mckenzie2021}, the degree of self-enrichment is strongly sensitive to the star formation and feedback prescriptions adopted in the simulation. For example, the sole thermal feedback from supernovae implemented in their simulations is not effective in removing gas from the cloud, allowing star formation to proceed for a long time. This results in enhanced self-enrichment within the cluster. Besides, by means of analytic arguments, \citet{Bland-Hawthorn+10} argue that the time over which star formation takes place is larger than the time required for the first supernovae to explode only in very massive clouds ($M > 10^7$~M$_\odot$ under the typical conditions of high-redshift environments, where GCs were born). High-resolution simulations able to capture the details of star formation and feedback are clearly required to shed light on this scenario.

Another interesting outcome of this work is that, both in NGC\,6362 and NGC\,6838, the metallicity spread among 2G stars is generally lower than the metallicity spread among 1G stars. This result would provide a challenge for scenarios of multiple population formation based on accretion of material processed in massive stars onto protostars.
\citet{gieles2018a} suggest that supermassive stars (SMSs) with masses bigger than $\sim 10^{3}$~M$_\odot$ are responsible for the distinctive chemical composition of 2G stars in GCs \citep[see also][]{denissenkov2014a}. In this scenario, protostars accrete the hot-hydrogen burning material processed in SMSs and expelled through strong SMS winds. The processed material, diluted with original gas, is responsible for the observed chemical composition of 2G stars. Similarly, in the scenario proposed by \citet{bastian2013a}, pre-MS stars are polluted with processed material released by interacting massive binaries and fast-rotating stars. The released gas is first accreted into the circumstellar discs and then onto the pre-MS stars, thus addressing the origin of the HeCNONa abundance anomalies of 2G stars \citep[see][for critical discussion on the formation scenarios]{renzini2015a}. 
More recently, \citet[][]{wang2020a} investigated the role of mergers of binary star
components for the formation of multiple stellar populations and suggested that 2G stars form from the gas polluted by mergers-driven
ejection winds. 

In these scenarios, 1G and 2G stars are expected to share the same iron abundance. In particular, an iron spread among 1G stars should result in nearly identical iron variations among the 2G. This fact is in contrast with what is observed in both NGC\,6362 and NGC\,6838, where the 2G exhibits small, if any, internal iron variations when compared with the 1G.

Alternative scenarios suggest that GCs experienced multiple star-formation episodes. 1G stars are supposed to form first. Their formation stops when feedback from massive stars clears any remaining gas from the cluster. At a later time, 2G stars form in the innermost and densest cluster region from original gas (i.e., gas that shares the same abundances as the 1G stars) reaccreted from the ISM and diluted with the material processed by more-massive 1G stars \citep[e.g.,\,][]{ventura2001a, dantona2004a, decressin2007a, dercole2008a, demink2009a, dercole2010a, denissenkov2014a,  dantona2016a, calura2019a}. Even though the level of iron inhomogeneties present in the ISM gas is expected to be the same, the lower metallicity spread observed among 2G stars can be explained by a different dynamical state of the proto-cluster environment. We know that the time required to smooth out initial chemical inhomogeneities in the gas is smaller in denser environments characterized by a higher gas velocity dispersion (see above). Hence, the chemical homogeneity of 2G stars is consistent with the idea that they formed in highly dense environments, as predicted by multi-generation scenarios. 

In this context, the evidence that 1G stars of M\,3- and M\,13-like GCs exhibit different internal iron variations may indicate different properties of the formation environment. Specifically, the smaller iron variation of M\,13-like GCs is consistent with a scenario where these clusters formed in high-density environments as suggested by  \citet[]{tailo2019a, tailo2019b, tailo2020, tailo2021a} based on the larger amount of RGB mass loss of 1G stars in these clusters.

\section*{acknowledgments} 
\small
We thank the anonymous referee for the useful suggestions. This work has received funding from the European Research Council (ERC) under the European Union's Horizon 2020 research innovation programme (Grant Agreement ERC-StG 2016, No 716082 'GALFOR', PI: Milone, http://progetti.dfa.unipd.it/GALFOR).
APM, MT, and GC acknowledge support from MIUR through the FARE project R164RM93XW SEMPLICE (PI: Milone). APM, ED, and MC acknowledge support from the PRIN-MIUR program 2017Z2HSMF (PI: Bedin).

\section*{Data Availability}
The data underlying this article will be shared on reasonable request to the corresponding author.

\begin{table*}
    \centering
    \caption{Internal iron variations among 1G stars ($\rm \delta [Fe/H]_{\rm 1G}$) for the 55 Galactic GCs studied in this work.}
    \begingroup
    \setlength{\tabcolsep}{10pt} 
    \renewcommand{\arraystretch}{1.} 
    \begin{tabular}{cccc}
    \hline \\[-.3cm]
         Cluster ID & $\rm \delta [Fe/H]_{\rm 1G}$ & Cluster ID & $\rm \delta [Fe/H]_{\rm 1G}$  \\[.1cm]
         \hline \\[-.4cm]
         NGC\,0104 & $0.087\pm0.009$ & NGC\,6304 & $0.125\pm0.028$ \\
         NGC\,0288 & $0.066\pm0.007$ & NGC\,6341 & $0.156\pm0.022$ \\
         NGC\,0362 & $0.075\pm0.010$ & NGC\,6352 & $0.064\pm0.018$ \\
         NGC\,1261 & $0.129\pm0.022$ & NGC\,6362 & $0.051\pm0.020$ \\
         NGC\,1851 & $0.063\pm0.007$ & NGC\,6366 & $0.017\pm0.030$ \\
         NGC\,2298 & $0.234\pm0.046$ & NGC\,6397 & $0.130\pm0.019$ \\
         NGC\,2808 & $0.110\pm0.010$ & NGC\,6441 & $0.074\pm0.007$ \\
         NGC\,3201 & $0.142\pm0.038$ & NGC\,6496 & $0.080\pm0.011$ \\
         NGC\,4590 & $0.123\pm0.015$ & NGC\,6535 & $0.128\pm0.022$ \\
         NGC\,4833 & $0.238\pm0.023$ & NGC\,6541 & $0.125\pm0.014$ \\
         NGC\,5024 & $0.307\pm0.029$ & NGC\,6584 & $0.121\pm0.028$ \\
         NGC\,5053 & $0.096\pm0.024$ & NGC\,6624 & $0.112\pm0.016$ \\
         NGC\,5139 & $0.148\pm0.011$ & NGC\,6637 & $0.061\pm0.009$ \\
         NGC\,5272 & $0.252\pm0.015$ & NGC\,6652 & $0.087\pm0.011$ \\
         NGC\,5286 & $0.172\pm0.012$ & NGC\,6656 & $0.195\pm0.038$ \\
         NGC\,5466 & $0.091\pm0.055$ & NGC\,6681 & $0.056\pm0.012$ \\
         NGC\,5897 & $0.129\pm0.030$ & NGC\,6717 & $0.019\pm0.010$ \\
         NGC\,5904 & $0.135\pm0.027$ & NGC\,6723 & $0.111\pm0.011$ \\
         NGC\,5927 & $0.215\pm0.023$ & NGC\,6752 & $0.106\pm0.017$ \\
         NGC\,5986 & $0.074\pm0.006$ & NGC\,6779 & $0.192\pm0.083$ \\
         NGC\,6093 & $0.107\pm0.010$ & NGC\,6809 & $0.115\pm0.011$ \\
         NGC\,6101 & $0.079\pm0.016$ & NGC\,6838 & $0.067\pm0.010$ \\
         NGC\,6121 & $0.039\pm0.031$ & NGC\,6934 & $0.123\pm0.028$ \\
         NGC\,6144 & $0.161\pm0.031$ & NGC\,6981 & $0.151\pm0.028$ \\
         NGC\,6171 & $0.066\pm0.011$ & NGC\,7078 & $0.204\pm0.014$ \\
         NGC\,6205 & $0.113\pm0.024$ & NGC\,7089 & $0.156\pm0.023$ \\
         NGC\,6218 & $0.060\pm0.015$ & NGC\,7099 & $0.000\pm0.020$ \\
         NGC\,6254 & $0.173\pm0.022$ & & \\[.1cm]
         \hline \hline
    \end{tabular}
    \endgroup     
    \label{Fe_table}
\end{table*}
\begin{table*}
    \centering
    \caption{Relation between the internal iron variations among 1G stars ($\rm \delta [Fe/H]_{\rm 1G}$), obtained in this work, and 20 GC parameters. For each parameter the Spearman's rank correlation coefficient is reported together with the associated p-value, indicative of the significance of the relation, and the number of degrees of freedom.}
    \begingroup
    \setlength{\tabcolsep}{6pt} 
    \renewcommand{\arraystretch}{1.2} 
    \begin{tabular}{cccc}
    \hline \\[-.5cm]
        Parameter &  $\rm \delta [Fe/H]_{\rm 1G}$ & Parameter & $\rm \delta [Fe/H]_{\rm 1G}$  \\[.1cm]
         \hline \\[-.5cm] 
         Age (MF09) & $-0.16,0.247,52$ & $f_{\rm bin}^{\rm C}$ & $0.44,0.008,33$ \\
         Age (D10)  & $0.23,0.098,53$ & $f_{\rm bin}^{\rm C-HM}$  & $0.25,0.094,44$ \\
         Age (V13)  & $0.15,0.286,48$ & $f_{\rm bin}^{\rm oHM}$  & $0.05,0.742,39$  \\ 
         Age (T20)  & $0.11,0.476,44$ & \emph{c}  & $0.06,0.656,53$ \\
         $\rm [Fe/H]$  & $-0.43,<0.001,53$ & $\log(\rm \tau_{hm})$  & $0.31,0.020,53$ \\
         $\log(\rm M/M_{\odot})$ & $0.35,0.008,53$ & $\mu_{V}$  & $0.02,0.884,53$  \\
         $\log(\rm M_{in}/M_{\odot})$ & $0.17,0.213,53$ & $\log(\rho_{\rm c})$  & $-0.08,0.547,53$  \\
         $\log(\rm M_{1G}/M_{\odot})$ & $0.39,0.004,50$ & $\sigma_{\rm v}$  & $0.19,0.174,53$ \\
         $\rm M_{V}$ & $-0.36,0.007,53$ & $\rm v_{\rm esc}$ & $0.19,0.175,53$ \\
         $\epsilon$  & $0.23,0.091,51$ & $\rho_{0}$  & $0.06,0.658,53$ \\[.1cm]
         \hline \hline
    \end{tabular}
    \endgroup
    \label{correlations_table}
\end{table*}

\bibliography{ms}{}

\label{lastpage}
\end{document}